\documentclass{entcs} 
\usepackage{tikz}
\usetikzlibrary{arrows,backgrounds,shapes}
\usepackage{hyperref}
\usepackage{listings}
\usepackage{latexsym, amssymb, amsmath, amsthm, amsfonts, yfonts, bbm, mathrsfs}
\usepackage[bbgreekl]{mathbbol}
\usepackage{stmaryrd,graphics,color}
\usepackage{subfigure}
\usepackage{algorithm,algpseudocode}

\def\lex{\mathtt{Lex}}

\def\parser{\mathtt{Parser}}

\def\token{\mathtt{Tokens}}

\def\fa{\mathsf{FA}_{/\equiv}}
\def\Inter{\mathsf{Interval}}
\def\Bool{\mathsf{Bool}}
\newcommand{\pp}[1]{ ^{#1.}}
\def\reflect{\mbox{\bf reflect}}
\def\rand{\mbox{\bf rand}}

\def\whilec{\mbox{\bf while}}
\def\ifc{\mbox{\bf if}}
\def\skipc{\mbox{\bf skip}}
\def\wid{\triangledown}
\def\pun{\mathfrak{P}}
\def\bp{\texttt{p}}
\def\regf{\mathtt{Regex}}

\usepackage{stmaryrd}
\def\ie{\emph{i.e.,\ }}

\def\Comm{{\tt c}}
\def\P{\tt P}
\def\Q{\tt Q}
\def\prog{\tt P}
\newcommand{\pc}[1]{\Loc_{#1}}
\newcommand{\pcf}[1]{\mbox{\sf Pl$_{\tiny #1}$}}
\newcommand{\stm}[1]{\mbox{\sf Stm$_{\tiny #1}$}}
\def\Exp{{\tt e}}
\def\Aexp{{\tt a}}
\def\rel{\leadsto}
\def\Bexp{{\tt b}}
\def\Sexp{{\tt s}}
\def\Rexp{{\tt r}}
\def\Dexp{{\tt d}}

\DeclareMathOperator{\AexpS}{AExp}
\DeclareMathOperator{\BexpS}{BExp}
\DeclareMathOperator{\ExpS}{Exp}
\DeclareMathOperator{\SexpS}{SExp}
\DeclareMathOperator{\Ccomms}{Comm}
\DeclareMathOperator{\CommS}{\textsf{DImp}}
\DeclareMathOperator{\Loc}{\textsf{PLines}}
\DeclareMathOperator{\Conf}{\mathbb{C}}
\DeclareMathOperator{\Statec}{\mathbb{c}}
\DeclareMathOperator{\Store}{\mathbb{S}}

\DeclareMathOperator{\mem}{\mathbb{m}}
\DeclareMathOperator{\amem}{\mathbb{m}^\sharp}
\DeclareMathOperator{\store}{\mathbb{s}}
\DeclareMathOperator{\tstore}{\mathbb{t}}
\DeclareMathOperator{\astore}{\mathbb{s}^\sharp}

\def\grasse#1{{\llbracket #1 \rrbracket}}
\def\grasstr#1{\langle\!|#1|\!\rangle}
\DeclareMathOperator{\true}{\textbf{tt}}
\DeclareMathOperator{\false}{\textbf{ff}}
\def\grasseb#1{{\llparenthesis\hspace*{0.2ex} #1 \hspace*{0.2ex}\rrparenthesis}}

\def\int{\mathit{int}}

\newcommand{\trad}[1]{\lbag #1\rbag}

\newcommand{\li}{\ar@{-}} 
\newcommand{\lp}{\ar@{.}} 
\newcommand{\fp}{\ar@{.>}} 

\def\bbbz{{\mathbb{Z}}}

\newcommand{\ccL}{\mathscr{L}}

\newcommand{\Int}{\ensuremath{\mathsf{Int}}}
\newcommand{\Exe}{\ensuremath{\mathsf{Exe}}}

\newcommand{\id}{\ensuremath{\mathit{id}}}

\newcommand{\Var}{\ensuremath{\textsf{Var}}}
\newcommand{\avalue}{\ensuremath{\textsf{AbstVal}}}
\newcommand{\vars}{\ensuremath{\textsf{vars}}}
\newcommand{\ra}{\rightarrow}
\newcommand{\la}{\leftarrow}
\newcommand{\Ra}{\Rightarrow}

\newcommand{\Lra}{\Leftrightarrow}
\newcommand{\lra}{\longrightarrow}

\newcommand{\ov}{\overline}

\def\defi{\triangleq}

\def\fT{{\mathfrak{T}}}

\newcommand{\integer}{\mathbb{Z}}

\def\ok#1{\mbox{\raisebox{0ex}[1ex][1ex]{$#1$}}}

\newcommand{\comment} [1]{}

\def\lfp{{\sf lfp\/}}

\def\rarr#1{\mbox{\raisebox{0ex}[1ex][1ex]{$
  \mathrel{\mathop{
\hspace*{1pt}\longrightarrow\hspace*{1pt}}\limits^{\,_{\mbox{\tiny 
\hspace*{-2.2pt}#1}}}}$}}} 

\def\grass#1{\llbracket#1\rrbracket}

\def\defemb#1#2{\expandafter\def\csname #1\endcsname
                              {\relax\ifmmode #2\else\hbox{$#2$}\fi}}
 
\def\2c-math#1#2{{\par\medskip\noindent ${#1}$
                      \par\smallskip
                        \noindent\hspace*{\fill} ${#2}$}
                           \\[10pt]}
\def\cAA{\ensuremath{A}}

\def\bB{\mathbb{B}}                  
\defemb{aF}{{\cal F}^\cAA}
\defemb{sF}{{\cal F}^{\star}}

\def\ba{\mathbf{a}}

\def\bbf{\mathbf{f}}

\def\bb{\mathbf{b}}

\defemb{cB}{{\cal B}}
\defemb{cC}{{\cal C}}
\defemb{cD}{{\cal D}}
\defemb{cE}{{\cal E}}
\defemb{cF}{{\cal F}}
\defemb{cG}{{\cal G}}
\defemb{cH}{{\cal H}}
\defemb{cI}{{\cal I}}
\defemb{cJ}{{\cal J}}
\defemb{cK}{{\cal K}}
\defemb{cL}{\ccL}
\defemb{cM}{{\cal M}}
\defemb{cN}{{\cal N}}
\defemb{cO}{{\cal O}}
\defemb{cP}{{\cal P}}
\defemb{cQ}{{\cal Q}}
\defemb{cR}{{\cal R}}
\defemb{cS}{{\cal S}}
\defemb{cT}{{\cal T}}
\defemb{cU}{{\cal U}}
\defemb{cV}{{\cal V}}
\defemb{cW}{{\cal W}}
\defemb{cY}{{\cal Y}}
\defemb{cX}{{\cal X}}
\defemb{cZ}{{\cal Z}}
\defemb{cHys}{{\cH_{\cY}^{\cS}}}
\defemb{cHy}{{\cH_{\cY}}}
\defemb{ld}{\dashv}
\defemb{rd}{\vdash}
\defemb{cGH}{{\cal GH}}
\defemb{cGWP}{{\cal GWP}}
\defemb{tL}{{\tt L}}
\defemb{tH}{{\tt H}}
\defemb{tT}{{\tt T}}
\defemb{tS}{{\tt S}}

\def\lin{\mbox{\cL$_{\cI}$}}
\def\lout{\mbox{\cL$_{\cO}$}}

\def\oout{\mbox{\sc fa$_{\cO}$}}

\newcommand{\sset}[2]{\{#1  ~|~ #2\}}
\newcommand{\labb}[1]{\textsf{lab}(#1)}

\newcommand{\ssetf}[2]{\left\{#1  \left |
                               \begin{array}{l}#2\end{array}
                          \right.     \right\}}

\newcommand{\set}[1]{\left\{
                         \begin{array}{l}#1
                         \end{array}
                     \right\}}

\newcommand{\conc}[1]{\tau^{\dot{#1}\:\frown}}

\newcommand{\nat}{\mathbb{N}}

\def\tuple#1{\langle #1 \rangle}

\newtheorem{theorem}{Theorem}[section]

\newtheorem{example}[theorem]{Example}

\def\nil{\:'\;'}
\newcommand{\mstr}[1]{\:'\!#1'}
\newcommand{\subst}[3]{\textbf{substr}(#1,#2,#3)}
\def\length{\textbf{len}}
\def\num{\textbf{num}}
\def\conc{\centerdot}
\def\sep{\mbox{\tt \$}}
\def\Mem{\mathbb{M}}

\def\lastname{Arceri \& Dalla Preda \& Giacobazzi \& Mastroeni}

\begin{document}
  \begin{frontmatter}
\title{SEA: String Executability Analysis by Abstract Interpretation}

\author{Vincenzo Arceri\thanksref{myemail}} 
  \thanks[myemail]{Email: \href{mailto:vincenzo.arceri@univr.it} {\texttt{\normalshape vincenzo.arceri@univr.it}}} 
\author{Mila Dalla Preda\thanksref{mdpemail}} 
  \thanks[mdpemail]{Email: \href{mailto:mila.dallapreda@univr.it} {\texttt{\normalshape mila.dallapreda@univr.it}}}
\author{Roberto Giacobazzi\thanksref{rgemail}} 
  \thanks[rgemail]{Email: \href{mailto:roberto.giacobazzi@univr.it} {\texttt{\normalshape roberto.giacobazzi@univr.it}}}
\author{Isabella Mastroeni\thanksref{imemail}} 
  \thanks[imemail]{Email: \href{mailto:isabella.mastroeni@univr.it} {\texttt{\normalshape isabella.mastroeni@univr.it}}}
  \address{Department of Computer Science, University of Verona, Italy}

\begin{abstract}
Dynamic languages often employ reflection primitives to turn dynamically generated text into executable code at run-time. These features make standard static analysis extremely hard if not impossible because its essential data structures, i.e., the control-flow graph and the system of recursive equations associated with the program to analyse, are themselves dynamically mutating objects. 
We introduce SEA, an abstract interpreter for automatic sound string executability analysis of dynamic languages employing bounded (i.e, finitely nested) reflection and dynamic code generation. 
Strings are statically approximated in an abstract domain of finite state automata with basic operations implemented as symbolic transducers. SEA combines standard program analysis together with string executability analysis. The analysis of a call to reflection determines a call to the same abstract interpreter over a code which is synthesised directly from the result of the static string executability analysis at that program point. The use of regular languages for approximating dynamically generated code structures allows SEA to soundly approximate safety properties of self modifying programs yet maintaining efficiency.
Soundness here means that the semantics of the code synthesised by the analyser to resolve reflection over-approximates the semantics of the code dynamically built at run-rime by the program at that point. 
\end{abstract}
\end{frontmatter}

%

\section{Introduction} 

\paragraph*{Motivation.}
The possibility of dynamically build code instructions as the result of text manipulation is a key aspect in dynamic programming languages. With reflection, programs can turn text, which can be built at run-time, into executable code \cite{RichardsHBV11}. These features are often used in code protection and tamper resistant applications, employing camouflage for escaping attack or detection \cite{DMavrogiannopoulosKP11}, in malware, in mobile code, in web servers, in code compression, and in code optimisation, e.g., in Just-in-Time (JIT) compilers employing optimised run-time code generation.

While the use of dynamic code generation may simplify considerably the {\em art and performance of programming\/}, this practice is also highly dangerous, making the code prone to unexpected behaviour and malicious exploits of its dynamic vulnerabilities, such as code/object-injection attacks for privilege escalation, data-base corruption, and malware propagation. It is clear that more advanced and secure functionalities based on reflection could be permitted if we better master how to safely generate, analyse, debug, and deploy programs that dynamically generate and manipulate code.

\begin{figure}
\begin{center}
\includegraphics[scale=.7]{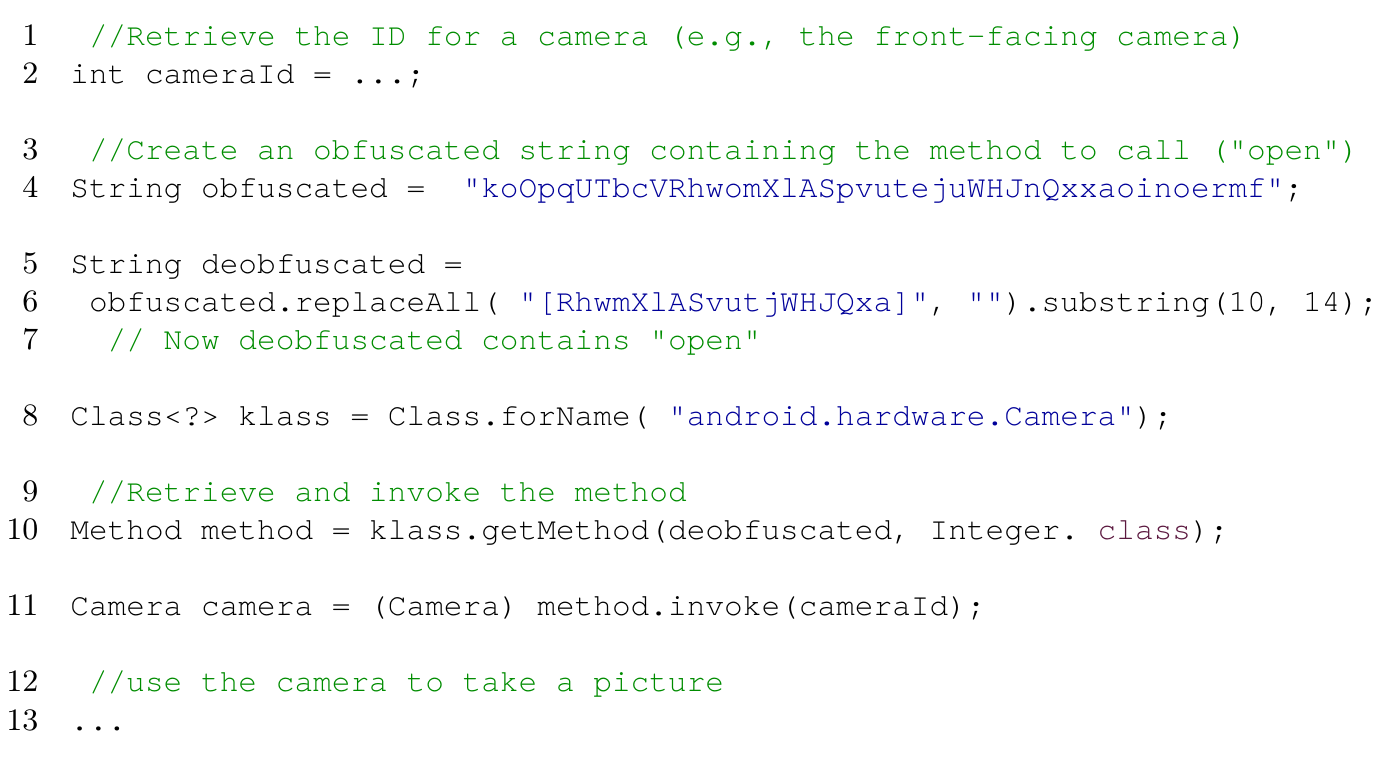}
\end{center}
\caption{A template rasomware obfuscated attack}\label{rasom}
\end{figure}

There are lots of good reasons to analyse when a program builds strings that can be later executed as code. Consider the code in Fig.~\ref{rasom}. This is a template of a ransomware that calls a method (``{\tt open}'') by manipulating an obfuscated string which is built in the code. Analysing the flow of the strings corresponds here to approximate the set of strings that may be turned into code at run-time. This possibility would provide important benefits in the analysis of dynamic languages, without ignoring reflection, in automated deobfuscation of dynamic obfuscators, and in the analysis of code injection and XSS attacks. 

\paragraph*{The problem.}
A major problem in dynamic code generation is that static analysis becomes extremely hard if not even impossible. This because program's essential data structures, such as the control-flow graph and the system of recursive equations associated with the program to analyse, are themselves dynamically mutating objects. In a sentence: {\em ''You can't check code you don’t see''\/} \cite{BesseyBCCFHHKME10}.

The standard way for treating dynamic code generation in programming is to prevent or even banish it, therefore restricting the expressivity of our development tools. Other approaches instead tries to combine static and dynamic analysis to predict the code structures dynamically generated \cite{Crispo2015,BoddenSSOM11}. Because of this difficulty, most analyses of dynamic languages do not consider reflection \cite{AnCFH11}, thus being inherently unsound for these languages, or implement ad-hoc or pattern driven transformations in order to remove reflection \cite{JensenJM12}. The design and implementation of a sound static analyser for self mutating programs is still nowadays an open challenge for static program analysis. 

\paragraph*{Contribution.}
In this paper we solve this problem by treating the code as any other dynamic structure that can be statically analysed by abstract interpretation \cite{CC77}.  We introduce SEA, a proof of concept for a fully automatic sound-by-construction abstract interpreter for string executability analysis of dynamic languages employing finitely nested (bounded) reflection and dynamic code generation. SEA carries a generic standard numerical analysis, in our case a well-known interval analysis, together with a new string executability analysis. Strings are approximated in an abstract domain of finite state automata (FA) with basic operations implemented as symbolic transducers and widening for enforcing termination. 

The idea in SEA is to re-factor reflection into a program whose semantics is a sound over-approximation of the semantics of the dynamically generated code. This allows us to continue with the standard analysis when the reflection is called on an argument that evaluates to code. In order to recognise whether approximated strings correspond to executable instructions, we approximate a parser as a symbolic transducer and, in case of success, we synthesise a program from the FA approximation of the computed string. 
The static analysis of reflection determines a call to the same abstract interpreter over the code synthesised from the result of the static string executability analysis at that program point. The choice of regular languages for approximating code structures provides efficient implementations of both the analysis and the code generation at analysis time. The synthesised program reflects most of the structures of the code dynamically generated, yet loosing some aspects such as the number of iterations of loops. 

Soundness here means that, if the approximated code extracted by the abstract interpreter is accepted by the parser, then the program may dynamically generate executable code at run-time. Moreover, because of the approximation of dynamically generated code structures in a regular language of instructions, any sound  and terminating abstract interpretation for safety (i.e., prefix closed) properties of the synthesised code, over-approximates the concrete semantics of the dynamically generated code. 
This means that a sound over-approximation of the concrete semantics of programs dynamically generating and executing code by reflection is possible for safety properties by combining string analysis and code synthesis in abstract interpretation. Even if nested reflection is not a common practice, the case of potentially unbound nested reflections, which may lead to non termination the analysis, can be handled as in \cite{JensenJM12} by fixing a maximal degree of nesting allowed. In this case, for programs exceeding the maximal degree of nested reflections, we may lose soundness. We briefly discuss how in SEA we may achieve an always sound analysis also for unbound reflection based on a widening with threshold over the recursive applications of the abstract interpreter. 

%
%

\section{Related Works} 
The analysis of strings is nowadays a relatively common practice in program analysis due to the widespread use of dynamic scripting languages. Examples of analyses for string manipulation are in \cite{DohKS09,ChristensenMS03,YuAB11,Thiemann05,Minamide05,imDS13}. 
The use of symbolic (grammar-based) objects in abstract domains is also not new (see \cite{CC95a,HeintzeJ94,Venet99}) and some works explicitly use transducers for string analysis in script sanitisation, see for instance \cite{HooimeijerLMSV11} and \cite{YuAB11}, all recognising that specifying the analysis in terms of abstract interpretation makes it suitable for being combined with other analyses, with a better potential in terms of tuning in accuracy and costs. None of these works use string analysis for analysing executability of dynamically generated code. 
In \cite{JensenJM12}, the authors introduce an automatic code rewriting techniques removing \texttt{eval} constructs in JavaScript applications. This work has been inspired by the work of Richards et al. \cite{RichardsHBV11} showing that \texttt{eval} is widely used, nevertheless in many cases its use can be simply replaced by JavaScript code without \texttt{eval}. 
In  \cite{JensenJM12}  the authors integrate a refactoring of the calls to \texttt{eval} into the TAJS data-flow analyzer. TAJS performs inter-procedural data-flow analysis on an abstract domain of objects capturing whether expressions evaluate to constant values. In this case \texttt{eval} calls can be replaced with an alternative code that does not use \texttt{eval}. It is clear that code refactoring is possible only when the abstract analysis recognises that the arguments of the \texttt{eval} call are constants. Moreover, they handle the presence of nested \texttt{eval} by fixing a maximal degree of nesting, but in practice they set this degree to $1$, since, as they claim, it is not often encountered in practice. The solution we propose allows us to go beyond constant values and refactor code also when then arguments of \texttt{eval} are elements of a regular language of strings. While this can be safely used for analysing safety properties of dynamically generated code, the use of our method for code refactoring has to take into account non-terminating code introduced by widening and regular language approximation. A more detailed comparison with TAJS is discussed in Sect.~\ref{evaluation}.

Static analysis for a static subset of PHP (i.e., ignoring {\tt eval}-like primitives) has been developed in \cite{BG09}. Static {\em taint analysis\/} keeping track of values derived from user inputs has been developed for self-modifying code by partial derivation of the Control-Flow-Graph \cite{WangJZL08}. The approach is limited to taint analysis, e.g., for limiting code-injection attacks. Staged information flow for JavaScript in \cite{ChughMJL09} with {\em holes\/} provides a conditional (a la abduction analysis in \cite{Giaco98}) static analysis of dynamically evaluated code. Symbolic execution-based static analyses have been developed for scripting languages, e.g., PHP, including primitives for code reflection, still at the price of introducing false negatives \cite{XieA06}. 

We are not aware of effective general purpose sound static analyses handling self-modifying code for high-level scripting languages. On the contrary, a huge effort was devoted to bring static type inference to object-oriented dynamic languages (e.g., see \cite{AnCFH11} for an account in Ruby) but with a different perspective: {\em Bring into dynamic languages the benefits of static ones -- well-typed programs don't go wrong\/}.
Our approach is different: {\em Bring into static analysis the possibility of handling dynamically mutating code}.
A similar approach is in \cite{AnckaertMB06} and \cite{PredaGD15} for binary executables. The idea is that of extracting a code representation which is descriptive enough to include most code mutations by a dynamic analysis, and then reform analysis on a linearization of this code.
On the semantics side, since the pioneering work on certifying self-modifying code in \cite{CSV07}, the approach to self-modifying code consists in treating machine instructions as regular mutable data structures, and to incorporate a logic dealing with code mutation within a la Hoare logics for program verification.  
TamiFlex \cite{BoddenSSOM11} also synthesises a program at every \texttt{eval} call by considering the code that has been executed during some (dynamically) observed execution traces. The static analysis can then proceed with the so obtained code without \texttt{eval}. It is sound only with respect to the considered execution traces, producing a warning otherwise.

\section{Preliminaries}\label{sft-sfa} 
\paragraph*{Mathematical Notation.} 
$S^*$ is the set of all finite sequences of elements in $S$. We often use bold letters to denote them. If $\mathbf{s} = s_1 \dots s_n \in S^\ast$, $s_i \in S$ is the $i$-th element and $|\mathbf{s}| \in \mathbb{N}$ its length. If $\mathbf{s}_1, \mathbf{s}_2 \in\Sigma^\ast$, $\mathbf{s}_1 \cdot \mathbf{s}_2\in\Sigma^\ast$ is their concatenation.

A set $L$ with ordering relation $\leq$ is a poset and it is denoted as $\tuple{L,\leq}$. 
Lattices $L$ with ordering $\leq$, least upper bound (lub) $\vee$, greatest lower bound (glb) $\wedge$, greatest element (top) $\top$, and least element (bottom) $\bot$ are denoted $\tuple{L,\leq,\vee,\wedge,\top,\bot}$. 
%
Given $f : S \rarr{} T$ and $g : T \rarr{} Q$ we denote with $g \circ f : S \rarr{} Q$  their composition, \ie $g \circ f = \lambda x.g(f(x))$. 
For $f,g : L \rarr{} D$ on complete lattices $f\sqcup g$ denotes the point-wise least upper bound, \ie $f\sqcup g = \lambda x.f(x)\vee g(x)$. 
$f$ is \emph{additive (co-additive)} if for any $Y \subseteq L, f(\vee_L Y ) = \vee_D f(Y)$ ($f(\wedge_L Y ) = \wedge_D f(Y ))$. The additive lift of a function $f:L\rarr{} D$ is the function  $\lambda X\subseteq L.\; \sset{f(x)}{x\in X}
\in\wp(L)\rarr{} \wp(D)$. We will often identify a function and its additive lift.
Continuity holds when $f$ preserves {\it lubs\/}'s of chains. 
For a continuous function $f$: $\lfp(f) = \bigwedge\sset{x}{x=f(x)}=\bigvee_{n\in\nat}f^n(\bot)$ where $f^0(\bot)=\bot$ and $f^{n+1}(\bot)=f(f^n(\bot))$. 

%
%
\paragraph*{Abstract Interpretation.}
Abstract interpretation establishes a correspondence between a concrete semantics and an approximated one called abstract semantics \cite{CC77,CC79}. 
In a Galois Connection (GC) framework, if $C$ and $A$ are complete lattices, a pair of monotone functions $\alpha: C \rarr{} A$ and $\gamma: A \rarr{} C$ forms 
a GC between $C$ and $A$ if for every $x \in C$ and $y \in A$ we have $\ok{\alpha(x) \leq_A y \Lra x \leq_C \gamma(y)}$. $\alpha$ (resp.\ $\gamma$) is the \emph{abstraction} (resp.\ \emph{concretisation}) and it is additive (resp.\ co-additive). Weaker forms of correspondence are possible, e.g., when $A$ is not a complete lattice or when only $\gamma$ exists.
%
In all cases, relative precision in $A$ is given by comparing the meaning of abstract objects in $C$, i.e., $x_1\leq_A x_2$ if $\gamma(x_1)\leq_C \gamma(x_2)$. 
%
%
If $\ok{f:C\rarr{}C}$ is a continuous function and $A$ is an abstraction of $C$ by means of the GC $\tuple{\alpha,\gamma}$, then 
$f$ always has a {\em best correct approximation\/} in $A$,  $\ok{f^A:A\lra A}$,  defined as
$\ok{f^{A} \defi \alpha \circ f \circ \gamma}$. Any approximation $\ok{f^\sharp:A\rarr{}A}$ of $f$ in $A$ is {\em sound\/} if $\ok{f^A\sqsubseteq f^\sharp}$. In this case we have the fix-point soundness $\ok{\alpha(\lfp f)\leq \lfp(f^A)\leq \lfp(f^\sharp)}$ (cf.\ \cite{CC77}). 
%
$A$ satisfies the ascending chain condition (ACC) if all ascending chains are finite. When $A$  is not ACC or when it lacks the limits of chains, convergence to the limit of the fix-point iterations can be ensured through widening operators. A \emph{widening operator} $\wid: A \times A \ra A$ approximates the lub, i.e., $\forall x,y \in A. x,y \leq_A (x \wid y)$ and it is such that for any increasing chain $x_1 \leq x_2 \leq \dots \leq x_n \leq \dots$ the increasing chain $w^0 = \bot$ and $w^{i+1} = w^i \wid x_i$ is finite.

\paragraph*{Finite State Automata (FA).}
A FA $A$ is a tuple $(Q,\delta,q_0,F, \Sigma)$, where $Q$ is the set of states, $\delta\subseteq Q\times  \Sigma \times Q$ is the transition relation, $q_0 \in Q$ is the initial state, $F \subseteq Q$ is the set of final states and $\Sigma$ is the finite alphabet of symbols. An element $(q,\sigma,q')\in\delta$ is called transition and is denoted $q' \in \delta(q,\sigma)$. Let $\omega \in  \Sigma^\ast$, $\hat{\delta}:Q \times  \Sigma^\ast \ra \wp(Q)$ is the transitive closure of $\delta$: $\hat{\delta}(q,\epsilon) = \{q\}$ and $\ok{\hat{\delta}(q,\omega \sigma) = \bigcup_{q'\in\hat{\delta}(q,\omega)}\delta(q',\sigma)}$. $\omega \in  \Sigma^\ast$ is accepted by $A$ if $\hat{\delta}(q_0,\omega) \cap F \neq \varnothing$. The set of all these strings defines the language $\cL(A)$ accepted by $A$. Given an FA $A$ and a partition $\pi$ over its states, we denote as $A/\pi = (Q',\delta',q_0',F', \Sigma)$ the \emph{quotient automaton} \cite{bookComp}.

\paragraph*{Symbolic Finite Transducers (SFT).}
We follow \cite{VeanesHLMB12} in the definition of SFTs and of their background structure.  
Consider a background universe $\cU_\tau$ of elements of type $\tau$, we denote with 
 $\bB$ to denote the elements of boolean type.
Terms and formulas are defined by induction over the background language and are well-typed. Terms of type $\bB$ are treated as formulas. $t:\tau$ denotes a term $t$ of type $\tau$, and $\mathit{FV}(t)$ denotes the set of its free variables. A term $t:\tau$ is \emph{closed} when  $\mathit{FV}(t) = \emptyset$. Closed terms have semantics $\grass{t}$. As usual $t[x/v]$ denotes the substitution of a variable $x:\tau$ with a term $v:\tau$. 
A $\lambda$-\emph{term} $f$ is an expression of the form $\lambda x . t$ where $x:\tau'$ is a variable  and $t:\tau''$ is a term such that $\mathit{FV}(t) \subseteq \{x\}$. The $\lambda$-term $f$ has type $\tau' \ra \tau''$ and its semantics is a function $\grass{f}: \cU_{\tau'} \ra \cU_{\tau''}$ that maps $a \in \cU_{\tau'}$ to $\grass{t[x/a]} \in \cU_{\tau''}$. 
Let $f$ and $g$ range over $\lambda$-terms. A $\lambda$-term of type $\tau \ra \bB$ is called a $\tau$-predicate.
Given a $\tau$-predicate $\varphi$, we write $a \in \grass{\varphi}$ for $\grass{\varphi}(a) = \mathit{true}$. Moreover,  $\grass{\varphi}$ can be seen as the subset of $\cU_{\tau}$ that satisfies $\varphi$. 
$\varphi$ is \emph{unsatisfiable} when $\grass{\varphi} = \emptyset$ and \emph{satisfiable} otherwise. 
%
A label theory \cite{VeanesHLMB12} for $\tau' \ra \tau''$ is associated with an effectively enumerable set of $\lambda$-terms of type $\tau' \ra \tau''$ and an effectively enumerable set of $\tau'$-predicates that is effectively closed under Boolean operations and relative difference, i.e., $\grass{\varphi \wedge \psi} = \grass{\varphi} \cap \grass{\psi}$, and $\grass{\neg\varphi} = \cU_{\tau'} \smallsetminus \grass{\varphi}$.
%
%
%
%
Let $\tau^\ast$ be the type of sequences of elements of type $\tau$. 
%
%
A Symbolic Finite Transducer \cite{VeanesHLMB12} (SFT) $T$ over $\tau' \ra \tau''$ is a tuple $T = \tuple{Q, q^0, F,R}$, where $Q$ is a finite set of states, $q^0 \in Q$ is the initial state, $F \subseteq Q$ is the set of final states and $R$ is a set of rules $(p,\varphi,\bbf,q)$ where $p,q \in Q$, $\varphi$ is a $\tau'$-predicate and $\bbf$ is a sequence of $\lambda$-terms over a given label theory for $\tau' \ra \tau''$.
%
A rule $(p,\varphi,\bbf,q)$ of an SFT $T$ is denoted as $p \stackrel{\varphi/\bbf}{\longrightarrow} q$.  
The sequence of $\lambda$-terms $\bbf:(\tau' \ra \tau'')^\ast$ can be treated as a function $\lambda x. [\bbf_0(x), \dots, \bbf_k(x)]$ where $k = |\bbf| - 1$. 
Concrete transitions are represented as rules. Let $p,q \in Q$, $a \in \cU_{\tau'}$ and $\bb \in \cU_{\tau''}^\ast$ then:
$\ok{
\stackrel{a/\bb}{\longrightarrow_T} q \; \Lra \; p \stackrel{\varphi/\bbf}{\longrightarrow_T} q \in R: a \in \grass{\varphi} \, \wedge \, \bb = \grass{\bbf}(a)
}$
%
Given two sequences $\ba \in \cU_{\tau'}^\ast$ and $\bb \in \cU_{\tau''}^\ast$, we write $q \stackrel{\ba/\bb}{\twoheadrightarrow} p$  when 
there exists a path of transitions from $q$ to $p$ in $T$ with input sequence $\ba=\ba_0\ba_1\cdots\ba_n$ and output sequence $\bb = \bb^0 \bb^1 \cdots \bb^n$, $n = |\ba| -1$ and $\bb^i$ denoting a subsequence of $\bb$, such that: 
$
\ok{
p = p_0 \stackrel{\ba_0/\bb^0}{\longrightarrow} p_1 \stackrel{\ba_1/\bb^1}{\longrightarrow} p_2 \dots p_n \stackrel{\ba_n/\bb^n}{\longrightarrow} p_{n+1} = q}
$.
SFT can have $\varepsilon$-transitions and they can be eliminated following a standard procedure. 
We assume 
$p\stackrel{\varepsilon/\varepsilon}{\longrightarrow}p$ for all $p \in Q$. 
%
The transduction of an SFT $T$ \cite{VeanesHLMB12}  over $\tau' \ra \tau''$ is a function $\fT_T:\cU_{\tau'}^\ast  \ra \wp(\cU_{\tau''}^\ast)$ where:
$\ok{
\fT_T(\ba) \defi \sset{\bb \in \cU_{\tau''}^\ast}{\exists q \in F: q^0 \stackrel{\ba/\bb}{\twoheadrightarrow} q} 
}
$

\paragraph*{SFT as FA Transformers.}
In the following, we will consider SFTs producing only one symbol in output for each symbol read in input. Namely, we consider SFTs with rules $(q,\varphi,f,q)$ where $f$ is a single $\lambda$-term of type $\tau' \ra \tau''$. Moreover, we consider SFTs and FA over finite alphabets, where the symbolic representation of SFT is useful for having more compact language transformers.  

In this section we show how, under these assumptions, SFTs can be seen as FA transformers. In particular, given an FA $A$ such that $\cL(A) \in \wp(\cU^\ast_{\tau'})$ and an SFT $T$ over $\tau' \ra \tau''$, we want to build the FA recognizing the language of strings in $\cU_{\tau''}^\ast$ obtained by modifying the strings in $\cL(A)$, according to the SFT $T$. To this end, we define the input language $\lin(T)$ of an SFT $T$ as the set of strings producing an output when processed by $T$, and the output language $\lout(T)$ as the set of strings generated by $T$. Formally: $\lin(T) \defi \sset{\ba \in \cU_{\tau'}^\ast}{\fT_T(\ba) \neq \emptyset}$ and  $\lout(T) \defi \{\bb \in \cU_{\tau''}^\ast ~|~ \bb \in \fT_T(\ba), \ba \in \lin(T)\}$. 

Consider $T =  \tuple{Q, q^0, F,R}$ over $\tau' \ra \tau''$, with $\cU_{\tau'}$ and $\cU_{\tau''}$ finite alphabets, and rules $(q,\varphi,f,p) \in R$ such that $f$ are $\lambda$-terms of type $\tau' \ra \tau''$. According to \cite{DMG-sas16} it is possible to build an FA $\oout(T)$ recognising the output language of $T$, i.e., $\cL(\oout(T)) = \lout(T)$. In particular, $\oout(T) \defi (Q,\delta,q_0,F,\cU_{\tau''})$ where $\delta = \{(q,b,p)~|~(q,\varphi,f,p) \in R, b \in \grass{f(\varphi)}\}$. Observe that $\grass{f(\varphi)}$ is finite since $\varphi$ is a predicate over a finite alphabet.
%
We can associate an SFT $\cT(A)$ to an FA $A$, where the input and output languages of $\cT(A)$ are the ones recognized by the FA $A$. Formally, given an FA $A=(Q,\delta,q_0,F,\cU_\tau)$, we define the output SFT over $\tau \ra \tau$ as $\cT(A) \defi \tuple{Q,q_0,F,R^\texttt{id}}$ where $R^\texttt{id} \defi \sset{(p,\sigma,\id,q)}{(p,\sigma,q)\in \delta}$\footnote{We denote by $\sigma$ the predicate requiring the symbol to be equal to $\sigma$.}
%
and the transduction is:
\[
\fT_{\cT(A)}(\ba) = 
\left\{
\begin{array}{ll}
\ba & \mbox{if } \ba \in \cL(A)\\
\emptyset & \mbox{otherwise}
\end{array}
\right. 
\]
These definitions allow us to associate FAs with SFTs and vice-versa and. 
According to \cite{VeanesHLMB12}, we define the composition of two transductions $\fT_1$ and $\fT_2$ as:
$$
\fT_1 \diamond \fT_2 \defi \lambda \bb . \bigcup _{\ba \in \fT_1(\bb)}\fT_2(\ba)
$$
Observe that the composition $\diamond$ applies first $\fT_1$ and then $\fT_2$.
%
It has been proved that if $T_1$ and $T_2$ are SFTs over composable label theories, then there exists an SFT $T_1 \diamond T_2$ that is obtained effectively from $T_1$ and $T_2$ such that $\fT_{T_1\diamond T_2} = \fT_1 \diamond \fT_2$  (see \cite{VeanesHLMB12,TR-bek} for details). 
%
At this point, given an FA $A$ with $\cL(A) \in \wp(\cU_{\tau'}^\ast)$ and an SFT $T$ over $\tau' \ra \tau''$, we can model the application of $T$ to 
$\cL(A)$ as the composition $\cT(A) \diamond T$ where
the language recognized by the FA $A$ becomes the input language of the SFT $T$. $\fT_{\cT(A) \diamond T}=\fT_T(\bb)$ if $\bb \in \cL(A)$, it is $\emptyset$ otherwise.
%
Observe that, the FA recognizing the output language of $\cT(A) \diamond T$ is the FA obtained by transforming $A$ with $T$. Indeed, $\cL(\oout(\cT(A) \diamond T)) = \{\bb \in \Gamma^\ast ~|~ \bb \in \fT_T(\ba), \ba \in \cL(A)\}$.
Thus, we can say that an SFT $T$ transforms an FA $A$ into the FA $\oout(\cT(A) \diamond T)$.


\section{A Core Dynamic Programming Language }\label{sect:dimp} 
\subsection{The dynamic language}
We introduce a core imperative deterministic dynamic language $\CommS$, in the style of {\sc Imp} for its imperative fragment and of dynamic languages, such as PHP or JavaScript, as far as string manipulation is concerned, with basic types integers in $\bbbz$, booleans, and strings of symbols over a finite alphabet $\Sigma$. Programs $\prog$ are labeled commands in $\CommS$ built as in Figure~\ref{synta}, on a set of variables $\Var$ and line of code labels $\pc{\prog}$ with typical elements $l\in\pc{\prog}$. 
\begin{figure}
{\footnotesize
  \begin{align*}
  \ExpS\ni \Exp &::= \; \Aexp\mid\Bexp\mid \Sexp\\
   \AexpS \ni \Aexp  &::= \; x \mid n \mid \rand() \mid \length(\Sexp)\mid \num(\Sexp) \mid\\ 
   & \Aexp + \Aexp \mid \Aexp -\Aexp \mid \Aexp*\Aexp\qquad(\mbox{where}\ n\in\integer)\\
    \BexpS \ni \Bexp &::= \; x \mid\true\mid\false \mid \Exp=\Exp \mid 
    \Exp > \Exp \mid \Exp < \Exp \mid \Bexp\wedge\Bexp \mid \neg {\tt b} \\
    \SexpS\ni \Sexp &::= \; x  \mid
    \nil \,\mid \mstr{\sigma} \mid \Sexp\conc\sigma\mid
   \subst{\Sexp}{\Aexp}{\Aexp}\\
   & (\mbox{where}\ \sigma\in\Sigma)\\
     \Ccomms\ni\Comm &::= \;\skipc;\mid x:=\Exp;\mid \Comm\Comm\mid \textbf{if} ~{\tt b}~\{\Comm\}; \mid\\ 
    & \textbf{while} ~b~ \{\Comm\};\mid \reflect(\Sexp); \mid  x := \reflect(\Sexp); \\
    \CommS\ni \P &::=\Comm\sep\\
    \mathsf{Id}\ni x &\quad \mbox{Identifiers (strings not containing punctuation symbols)}
   \end{align*}}
   \caption{Syntax of $\CommS$}\label{synta}
\end{figure}
%
We assume that all terminal and non terminal symbols of $\CommS$ are in $\Sigma_{\CommS}\subseteq\Sigma^*$. Thus, the language recognized by the context free grammar (CFG) of $\CommS$ is an element of $\wp((\Sigma_{\CommS})^\ast)$, i.e., $\CommS \subseteq (\Sigma_{\CommS})^\ast$. Given $\prog \in \CommS$ we associate with each statement a program line $l\in \pc{\prog}$. In order to simplify the presentation of the semantics, we suppose that any program is ended by a termination symbol $\sep$, labeled with the last program line denoted $l_e$. When a statement $\Comm$ belongs to a program $\prog$ we write $\Comm\in \prog$, then we define the auxiliary functions $\stm{\prog}: \pc{\prog}\ra \CommS$ be such that $\stm{\prog}(l)=\Comm$ if $\Comm$ is the statement in $\prog$ at program line $l$ (in the following denoted $\pp{l}\Comm$) and $\pcf{\prog}=\stm{\prog}^{-1}:\CommS\ra \pc{\prog}$ with the simple extension to blocks of $\prog$ instructions  $\pcf{\prog}(\Comm_1\Comm_2)=\pcf{\prog}(\Comm_1)$. 
In general, we denote by $\pcf{\prog}$ the set of all the program lines in $\P$\footnote{Note that, by definition a statement, or a block, $\Comm$ ends always with $;$.}.

Let $M\defi\Var \rarr{} \mathbb{Z} \cup \{\true,\false\} \cup \Sigma^*$ be the set of memory maps, ranged over by $m$, that assign values (integers, booleans or strings) to variables. 
$\grasseb{\Sexp}:M \rarr{} \Sigma^*$ denotes the semantics of string expressions. For strings $\Sexp_1,\Sexp_2\in\Sigma^*$, symbol $\delta\in\Sigma$ and values $n_1,n_2\in\bbbz$ we have that $\grasseb{\Sexp_1\conc\delta}m$ returns the concatenation of the string $\grasseb{\Sexp_1}m$ with the  symbol $\delta\in\Sigma$, i.e., $\grasseb{\Sexp_1}m\cdot\delta$. 
We abuse notation and use $\Sexp_1\conc\Sexp_2$ for string concatenation.  
The semantics $\grasseb{\subst{\Sexp}{n_1}{n_2}}m$ returns the sub-string of the string $\grasseb{\Sexp}m$ given by the $n_2$ consecutive symbols starting from the $n_1$-th one (we suppose $n_1\geq 0$)\footnote{The choice of considering only concatenation and substring derives form the fact that most of the operations on strings can be obtained as by using these two operations.}.
We denote with $\grasseb{\Aexp}: M \rarr{} \mathbb{Z}$ the semantics of arithmetic expressions where $\grasseb{\length(\Sexp)}m$ returns the length of the string $\grasseb{\Sexp}m$, and  $\grasseb{\num(\Sexp)}m$ returns the integer denoted by the string $\grasseb{\Sexp}m$ (suppose it returns the empty set if $\grasseb{\Sexp}m$ is not a number). The semantics of the other arithmetic expressions is defined as usual.  Analogously, $\grasseb{\Bexp}: M \ra \{\true,\false \}$  denotes the semantics of Boolean expressions where, given  $\Sexp_1,\Sexp_2\in\Sigma^*$, $\Sexp_2 <\Sexp_1$ is true iff $\Sexp_2\preceq\Sexp_1$ (prefix order). The semantics of the other Boolean expressions is defined as usual.

The update of memory $m$, for a variable $x$ with value $v$, is denoted $m[x/v]$.  
The semantics of $\reflect(\Sexp)$ evaluates the string $\Sexp$: if it is a program in $\CommS$ it executes it, otherwise the execution proceeds with the next command. Observe that $\Sexp \in \Sigma^\ast$ while $\CommS \in \wp((\Sigma_{\CommS})^\ast)$, for this reason we define $\ov{\CommS} \defi \{\ba \in \Sigma^\ast ~|~ \ba = \ba_1\cdot\ba_2 \cdot \ldots \cdot \ba_n, \ba_1\ba_2 \ldots  \ba_n \in \CommS\}$ as the set of sequences in $\Sigma^\ast$ that can be obtained by concatenating the sequences $\ba^i$ that act like symbols in a program in $\CommS$. We denote with $\ov{\Comm}$ the sequence of $\Sigma^\ast$ that corresponds to the sequence $\Comm \in (\Sigma_{\CommS})^\ast$.
At this point, before computing the semantics of $\ov{\Comm}$, we need to recognize which statements it denotes, building the corresponding string $\Comm \in \CommS$, and then to label this statements by using the function $\labb{\cdot}$, assigning an integer label to each statement in $\Comm\sep$ from $1$ to the final program point $l_e$. In the following, we say that $\Sexp$ evaluates to $\Comm$ when it assumes a value $\ov{\Comm} \in \ov{\CommS}$ corresponding to the concatenation of the sequences that are symbols of $\Comm$. 
The semantics of $x:=\reflect(\Sexp)$ evaluates expression $\Sexp$ and if it is $\Comm$ in $\CommS$ it proceeds by assigning $\nil$ to $x$ and executes $\Comm$, otherwise it behaves as a standard assignment. Formally, let $\Int:\CommS\times M\rarr{} M$ denote the semantics of programs, and $\grasseb{\cdot}m$ the evaluation of an expression in the memory $m$, then:

\vspace{-.2cm}
{\footnotesize
\begin{align*}
\Int(\pp{l}\reflect(\Sexp);\pp{l'}\Q, \textit{m})&=
\left \{
\begin{array}{ll}
\Int(\pp{l'}\Q,\textit{m'})\\
\qquad\mbox{if}\ \grasseb{\Sexp}m \cap \ov{\CommS} = \ov{\Comm}\ \wedge\\
\qquad \textit{m'}=\Int(\labb{\Comm},\textit{m})\\
\Int(\pp{l'}\Q,\textit{m})\quad \mbox{otherwise}
\end{array}
\right .
\end{align*}
\vspace{-.2cm}
}

We can observe that the way in which we treat the commands $\pp{l}\reflect(\Sexp)$ and $\pp{l}x:=\reflect(\Sexp)$ mimics the classical semantic model and implementation of reflection and reification as introduced in Smith \cite{Smith84}, see \cite{WandF88,DanvyM88} for details. 
In particular, when string $\Sexp$ evaluates to a program $\Comm$, the program control starts the execution of $\Comm$ before returning to the original code. The problem is that $\Comm$ may contain other $\reflect$ statements leading to the execution of new portions of code. Hence, each nested $\reflect$ is an invocation of the interpreter which can be seen as a new layer in the {\em tower} of interpretations: When a layer terminates the execution the control returns to the previous layer with the actual state.
Hence, the state $\Int(\pp{l}\reflect(\Sexp);\pp{l'}\Q,\textit{m})$, when string $\Sexp$ evaluates to a program $\Comm$, starts a new computation of $\labb{\Comm}$ from $m$. 
Once the execution of the tower derived from $\Comm$ terminates, the execution comes back to the continuation $\Q$, in the memory resulting from the execution of $\Comm$. It is known that in general the construction of the tower of interpreters may be infinite leading to a divergent semantics.
\begin{example}\label{infTower}
Consider the following program fragment $\P$:
\begin{equation*}
\pp{1}x:=\mstr{\reflect(x);\sep};\:\pp{2}\reflect(x);\:\pp{3}\sep
\end{equation*}
Suppose the initial memory is $m_\bot$ (associating the undefined value to each variable, in this case $x$). After the execution of the first assignment we have the memory $m_1=[x/\mstr{\reflect(x)}]$, on which we execute the $\reflect(x)$ statement. Since now, $\reflect(x)$ is executed starting from $m_1$, hence each $\reflect(x)$ activates a tower layer executing the statement in $x$, which is again $\reflect(x)$ starting from the same memory. Hence the tower has infinite height.
\end{example}


\subsection{Flow-sensitive Collecting Semantics}
\comment{
\begin{figure*}[ht]
{\footnotesize
\begin{tabular}{ll}
$\begin{array}{l}
\grasseb{\true}\mem \defi \true \\ 
\grasseb{\false}\mem \defi \false 
\end{array}
\quad 
\grasseb{\neg\Bexp}\mem \defi 
\left\{
\begin{array}{ll}
\true & \mbox{ if }  \grasseb{\Bexp}\mem = \false\\
\false & \mbox{ if } \grasseb{\Bexp}\mem = \true\\
\{\true,\false\} & \mbox{ otherwise}
\end{array}
\right.
$
&
$\grasseb{\Exp_1 = \Exp_2}\mem \defi 
\left\{
\begin{array}{ll}
\true & \mbox{ if } |\grasseb{\Exp_1}\mem| = |\grasseb{\Exp_2}\mem| = 1 \mbox{ and } \grasseb{\Exp_1}\mem = \grasseb{\Exp_2}\mem \\
\false & \mbox{ if } \grasseb{\Exp_1}\mem \cap \grasseb{\Exp_2}\mem = \varnothing \\
\{\true,\false\} & \mbox{ otherwise}
\end{array}
\right.
$\\ \ \\
$\grasseb{\Aexp_1 > \Aexp_2}\mem \defi 
\left\{
\begin{array}{ll}
\true & \mbox{ if } \forall x \in \grasseb{\Aexp_1}\mem, \forall y \in \grasseb{\Aexp_2}\mem: x > y\\
\false & \mbox{ if } \forall x \in \grasseb{\Aexp_1}\mem, \forall y \in \grasseb{\Aexp_2}\mem: x \leq y\\
\{\true,\false\} & \mbox{ otherwise}
\end{array}
\right.
$
&
$\grasseb{\Bexp_1 > \Bexp_2}\mem \defi 
\left\{
\begin{array}{ll}
\false & \mbox{ if }   \grasseb{\Bexp_2}\mem = \true\\
\true & \mbox{ if } \grasseb{\Bexp_1}\mem = \true \mbox{ or } \grasseb{\Bexp_2}\mem = \false\\
\{\true,\false\} & \mbox{ otherwise}
\end{array}
\right.
$\\ 
\mbox{(Analogous for $\grasseb{\Sexp_1>\Sexp_2}\mem$)} &\\
\end{tabular}}
\caption{Semantics of Boolean expression}\label{csem}
\end{figure*}}
Collecting semantics models program execution by computing, for each program point, the set of all the values that each variables may have. 
%
In order to deal with reflection we need to define an interpreter collecting values for each program point which, at each step of computation, keeps trace not only of the collection of values after the last executed program point $p$, but also of the values collected in all the other program points, both already executed and not executed yet. In other words, we define a flow sensitive semantics which, at the end of the computation, observes the trace of collections of values holding {\em at each program point}. In order to model this semantics, we model the concrete state not simply as a memory -- the current memory, but as the tuple of memories holding at each program point. It is clear that, at each step of computation, only the memory in the last executed program point will be modified. 
First, we define a collecting memory $\mem$, associating with each variable a set of values instead of a single value.
We define the set $\ok{\Mem\defi\Var \rarr{} \wp(\mathbb{Z}) \cup \Bool \cup \wp(\Sigma^*)}$ with meta-variable $\mem$, where $\Bool = \wp(\{\false,\true\})$. We define two particular memories, $\mem_\varnothing$ associating $\varnothing$ to any variable, and $\mem_\top$ associating the set of all possible values to each variable. 
The update of memory $\mem$ for a variable $x$ with set of values $v$ is denoted $\mem[x/v]$.
Finally, lub and glb of memories are $\mem_1\sqcup \mem_2(x)=\mem_1(x)\cup\mem_2(x)$ and $\mem_1\sqcap\mem_2(x)=\mem_1(x)\cap\mem_2(x)$.
\\ Then, in order to make the semantics flow-sensitive, we introduce a new notion of {\em flow-sensitive} store (in the following called store) $\Store \defi \pc{\prog}\rarr{}\Mem$ associating with each program line a memory.
We represent a store $\store \in \Store$ at a given line $l\in\pc{\prog}$ as a tuple $\tuple{x_1/v_{x_1},\ldots, x_n/v_{x_{n}}}$, where $v_{x_i}$ is the set of possible values of variable $x_i$. We use $\store_l$ to denote $\store(l)$, namely the memory at line $l$.  Given a store $\store$, the update of memory $\store_l$ with a new collecting memory $\mem$ is denoted $\store[\store_l \la \mem]$ and provides a new store $\store'$ such that $\store'_l=\store_l\sqcup\mem$ while $\forall l'\neq l$ we have $\store'_{l'}=\store_{l'}$.  

We abuse notation by denoting with $\grasseb{\cdot}$, not only the concrete, but also the collecting semantics of expressions, all defined as additive lift of the expression semantics. 
In particular, we denote by $\grasseb{\Bexp}^{\true}$ the maximal collecting memory making $\Bexp$ true, i.e., it is $\bigsqcup\sset{m\in M}{\grasseb{\Bexp}m=\true}\in\Mem$ (analogous for $\grasseb{\Bexp}^{\false}$). Hence, by $\mem\sqcap\grasseb{\Bexp}^{\true}$ we denote the memory $\mem'\defi\mem[x\in\vars(\Bexp)/\mem(x)\cap\grasseb{\Bexp}^{\true}(x)]$, where $\vars(\Bexp)$ is the set of variables of $\Bexp$. For instance if $\mem=[x/\{1,2,3\},y/\{1,2\}]$ and $\Bexp=(x<3)$, then $\grasseb{\Bexp}^{\true}=[x/\{1,2\},y/\top]$, hence $\mem\sqcap\grasseb{\Bexp}^{\true}=[x/\{1,2\},y/\{1,2\}]$. Finally, let $V\subseteq \Var$, by 
$\store_V$ we denote the store where for each program point $l$, the memory $\store_l$ is restricted only on the variables in $V$, by
$\lfp_V f(\store)$ we denote the computation of the fix point only on the variables $V$, i.e., we compute $\store$ such that $\store_V=f(\store)_V$.

We follow \cite{C00tcs} in the usual definition of the \emph{concrete collecting trace semantics} of a transition system $\tuple{\Conf,\rel}$ associated with programs in $\CommS$, where $\Conf=\CommS\times\Store$ is the set of states in the transition system with typical elements $\Statec\in\Conf$ and $\rel\subseteq\Conf\times\wp(\Conf)$ is a transition relation.
The state space in the transition system is the set of all pairs $\tuple{\Comm,\store}$ with $\Comm\in\CommS$ and $\store\in\Store$ representing the store computed by having executed the first statement in $\Comm$ and having its continuation still to execute. The transition relation generated by a program $\Comm \in\CommS$ is in Appendix. 
The axiom $\tuple{\pp{l_e}\sep,\store}$ identifies the final blocking states $\cB$.
When the next command to execute is $\pp{l}\reflect(\Sexp)$, we need to verify whether the evaluation of the string $\Sexp$ at program line $l$ returns a set of sequences of symbols of $\Sigma$ that contains sequences representing programs in $\CommS$. If this is the case, we proceed by executing the programs corresponding to $\grasseb{\tt s}\store_l$ with initial memory (at the first program line of $\Comm$) the memory holding at program line $l$, while the memories for all the other program points in $\Comm$ are initialized to $\mem_\varnothing$. 
%
When the next command to execute is an assignment of the form $\pp{l} x:= \reflect(\Sexp)$ we need to verify whether string  $\Sexp$ evaluates to a simple set of strings or to strings corresponding to programs in $\CommS$. If the evaluation of the strings in $\grasseb{\tt s}\store_l$ does not contain programs we proceed as for standard assignments. If $\grasseb{\tt s}\store_l$ returns programs in $\CommS$ then the assignment becomes an assignment of $\nil$ to variable $x$ and the execution of the programs corresponding to $\grasseb{\tt s}\store_l$.
Observe that, in order to verify whether the possible values assumed by a string $\Sexp$ at a program point $l$ are programs in $\CommS$, we check if the intersection $\grasseb{\Sexp}\store_l \cap \ov{\CommS}$ is not empty. Unfortunately, this step is in general undecidable, for this reason, in Section~\ref{approx2}, we provide a constructive methodology for deciding the executability of $\grasseb{\Sexp}\store_l$ and for synthesizing a program that can be executed in order to proceed with the analysis and obtain a sound result.    
The other rules model standard transitions. 

Given a program $\P\in\CommS$ and a set of initial stores $I$, we denote by $\cI\defi\sset{\Statec}{\Statec = \tuple{\P,\store}, \store \in I}$ the set of initial states.  
%
%
In sake of simplicity, we consider a partial collecting trace semantics observing only the finite prefixes of finite and infinite execution traces:
{\small \[
\cF(\P,\cI)=\ssetf{\Statec_0\Statec_1\ldots\Statec_n}{\Statec_0 \in \cI,\ 
\forall i<n.\; \Statec_i\rel\Statec_{i+1}}
\]  }\vspace{-.2cm}

It is known that $\cF(\P,\cI)$ expresses precisely invariant properties of program executions and it can be obtained by fix-point of the following trace set transformer $\mathtt{F}:\wp(\Conf^*)\rarr{}\wp(\Conf^*)$, starting form the set $\cI$ of initial configurations, such that $\cF(\P,\cI) = \lfp(\mathtt{F}_{\P,\cI})$.
{\small \[
\begin{array}{ll}
\mathtt{F}_{\P,\cI}\defi& \lambda X.\: \cI\:\cup 
                  \ssetf{\Statec_0\Statec_1\ldots\Statec_i\Statec_{i+1}}{\Statec_0\Statec_1\ldots\Statec_i\in X\!\!\\
\Statec_i\rel\Statec_{i+1}\!\!}
\end{array}
\]}

Finally, we can define the store projection of the partial collecting trace semantics (in the following simply called trace semantics) of a program $\P$ form an initial store $\store\in I$ as
{\small
\[
\grasstr{\P}\store\defi\ssetf{\store\store^1\ldots\store^n}{\exists \Statec_0\Statec_1\ldots\Statec_n\in \cF(\P,\{\store\}).\:\Statec_0=\tuple{\P,\store}\\ \wedge\ \forall i\in[1,n].\:\exists \P_i\in\CommS.\:\Statec_i=\tuple{\P_i,\store^i}}
\]}
\begin{example}\label{es1}
Consider the following $\CommS$ program $\P$ implementing an iterative count by dynamic code modification.
\begin{center}
{\small
\begin{tabular}{l}
$\pp{1}x:=1; \pp{2}\mathit{str}:= \mstr{\sep};$\\
$\pp{3}\whilec \; x < 3 \; \{\pp{4}\mathit{str} := \mstr{x:=x+1;} \conc\mathit{str};\:\pp{5}\reflect (\mathit{str});\};\:\pp{6}\sep$
\end{tabular}}
\end{center}
At each step of computation, let us denote by $\P$ the continuation of the program.
%
A portion of the iterative computation of the collecting semantics, starting from the store $\store^0$ such that, for each $l\in[1,6],\:\store^0_l=\mem_\varnothing$, is reported in Fig.~\ref{fig:es}. Note that, $\store_1=\mem_\varnothing$ at each step of computation, while $\store_2=[x/\{1\},\mathit{str}/\varnothing]$ after the execution of the first statement. Moreover, in sake of brevity, we will denote as $\mstr{s}$ the string $\mstr{x:=x+1}$.
\begin{figure*}[ht]
 \scalebox{0.7}[1]{%
 \vbox{%
\hspace{-.5cm}
{\tiny
\begin{tabular}{|l|l|l|l|l|l|}
\hline
 P & $\store_3$& $\store_4$& $\store_5$& $\store_6$\\
\hline\hline
$\pp{1}x:=1;\pp{2}\P$ &$\mem_\varnothing$&$\mem_\varnothing$&$\mem_\varnothing$&$\mem_\varnothing$\\
\hline
$\pp{2}\mathit{str}:=\nil; \pp{3}\P$  &$\mem_\varnothing$&$\mem_\varnothing$&$\mem_\varnothing$&$\mem_\varnothing$\\
\hline
$\pp{3}\whilec\ x<3\ \{\pp{4}\Comm\};\pp{6}\sep$ 
&$[x/\{1\},\mathit{str}/\{\nil\}]$&$\mem_\varnothing$&$\mem_\varnothing$&$\mem_\varnothing$\\
\hline
$\pp{4}str:= \mstr{x:=x+1;} \conc \mathit{str};\pp{5}\Comm_1; \pp{3}\P$ 
&$[x/\{1\},\mathit{str}/\{\nil\}]$&$[x/\{1\},\mathit{str}/\{\nil\}]$&$\mem_\varnothing$&$\mem_\varnothing$\\
\hline
{\color{red}$\pp{5}\reflect(\mathit{str});$}$ \pp{3}\P$ 
&$[x/\{1\},\mathit{str}/\{\nil\}]$&$[x/\{1\},\mathit{str}/\{\nil\}]$&$[x/\{1\},\mathit{str}/\{\nil,\mstr{s}\}]$&$\mem_\varnothing$\\
\hline
$\pp{3}\whilec\ x<3\ \{\pp{4}\Comm\};\pp{6}\sep$ &{\color{red}$[x/\{1,2\},\mathit{str}/\{\nil,\mstr{s}\}]$}&$[x/\{1\},\mathit{str}/\{\nil\}]$&$[x/\{1\},\mathit{str}/\{\nil,\mstr{s}\}]$&$\mem_\varnothing$\\
\hline
$\pp{4}str:= \mstr{x:=x+1;} \conc \mathit{str};\pp{5}\Comm; \pp{3}\P$ 
&$[x/\{1,2\},\mathit{str}/\{\nil,\mstr{s}\}]$&$[x/\{1,2\},\mathit{str}/\{\nil,\mstr{s}\}]$&$[x/\{1\},\mathit{str}/\{\nil,\mstr{s}\}]$&$\mem_\varnothing$\\
\hline
{\color{red}$\pp{5}\reflect(\mathit{str});$}$ \pp{3}\P$ 
&$[x/\{1,2\},\mathit{str}/\{\nil,\mstr{s}\}]$&$[x/\{1,2\},\mathit{str}/\{\nil,\mstr{s}\}]$&$[x/\{1,2\},\mathit{str}/\{\nil,\mstr{s},\mstr{s;s}\}]$&$\mem_\varnothing$\\
\hline
$\pp{3}\whilec\ x<3\ \{\pp{4}\Comm\};\pp{6}\sep$ &{\color{red} $[x/\{1,2,3,4\},\mathit{str}/\{\nil,\mstr{s},\mstr{s;s}\}]$}&$[x/\{1,2\},\mathit{str}/\{\nil,\mstr{s}\}]$&$[x/\{1,2\},\mathit{str}/\{\nil,\mstr{s},\mstr{s;s}\}]$&$\mem_\varnothing$\\
\hline
$\pp{6}\sep$ &$[x/\{1,2,3,4\},\mathit{str}/\{\nil,\mstr{s},\mstr{s;s}\}]$&$[x/\{1,2\},\mathit{str}/\{\nil,\mstr{s}\}]$&$[x/\{1,2\},\mathit{str}/\{\nil,\mstr{s},\mstr{s;s}\}]$&$[x/\{3,4\},\mathit{str}/\{\nil,\mstr{s},\mstr{s;s}\}]$\\
\hline
\end{tabular}
}
}
}
\caption{Iterative computation of the collecting semantics of program $\P$ in Example~\ref{es1}, with $s\defi x:=x+1$}\label{fig:es}

\end{figure*}
\noindent
With a different color, we highlight the execution of reflect activating a new analysis computation, and the memory $\store_3$ computed by the statements executed by the reflect. In particular, the first execution of $\mathbf{reflect}(\mathit{str})$ is such that $\grasseb{\mathit{str}}\store_5\cap\CommS=\{x:=x+1;\sep\}$. Moreover, the initial store $\store^\iota$ for the execution of $\reflect$ is such that $\store^\iota_1=\store_5$, and $\forall 1<l\leq l_e$ $\store_{l}=\mem_\varnothing$. $\labb{x:=x+1;\sep}=\pp{1}x:=x+1;\pp{2}\sep$, with $l_e=2$.
Now, the computation of $\labb{x:=x+1;\sep}$ is given in Fig.~\ref{fig:es2} on the left. In this case $\store^e_{l_e}=[x/\{2\},\mathit{str}/\{\nil,\mstr{s}\}]$, hence the new $\store_3$ is the least upper bound between this $\store^e_{l_e}$ and the previous $\store_3$, which is $[x/\{1,2\},\mathit{str}/\{\nil,\mstr{s}\}]$. The second time {\reflect} is executed, we have $\grasseb{\mathit{str}}\store_5\cap\CommS=\{x:=x+1';\sep, x:=x+1;x:=x+1;\sep\}$. The calling memory is $\store^{\iota}_1=\store_5=[x/\{1,2\},\mathit{str}/\{\nil,\mstr{s},\mstr{s;s}\}]$. Similarly to the previous case, the execution of $\pp{1}x:=x+1;\pp{2}\sep$ returns the least upper bound between $\store_3$ and $[x/\{2,3\}, \mathit{str}/\{\nil,\mstr{s},\mstr{s;s}\}]$ which is $[x/\{1,2,3\}, \mathit{str}/\{\nil,\mstr{s},\mstr{s;s}\}]$. Finally, the execution of $\pp{1}x:=x+1;\pp{2}x:=x+1;\pp{3}\sep$ is given in Fig.~\ref{fig:es2} (on the right). In this case, the resulting memory is the least upper bound between $\store_3$ and $[x/\{3,4\},\mathit{str}/\{\nil,\mstr{s},\mstr{s;s}\}]$, which is $[x/\{1,2,3,4\}, \mathit{str}/\{\nil,\mstr{s},\mstr{s;s}\}]$, which is also the least upper bound of all the resulting memories, i.e., the new $\store_3$.

\begin{figure*}[ht]
 \scalebox{0.7}[1]{%
 \vbox{%
\begin{center}
{\tiny
\begin{tabular}{|l|l|l||l|l|l|}
\hline
$\mathbf{reflect}(\mstr{s})$ & $\store_1$& $\store_2$&$\mathbf{reflect}(\mstr{s;s})$ &  $\store_2$&$\store_3$\\
\hline\hline
$\pp{1}x:=x+1;\pp{2}\sep$ &$[x/\{1\},\mathit{str}/\{\nil,\mstr{s}\}]$&$\mem_\varnothing$&
$\pp{1}x:=x+1;\pp{2}\P$&$\mem_\varnothing$&$\mem_{\varnothing}$\\
\hline
$\pp{2}\sep$  &$[x/\{1\},\mathit{str}/\{\nil,\mstr{s}\}]$&{\color{red}$[x/\{2\},\mathit{str}/\{\nil,\mstr{s}\}]$}&$\pp{2}x:=x+1;\pp{3}\sep$&$[x/\{2,3\},\mathit{str}/\{\nil,\mstr{s},\mstr{s;s}\}]$&$\mem_{\varnothing}$\\
\hline
&&&$\pp{3}\sep$&$[x/\{2,3\},\mathit{str}/\{\nil,\mstr{s},\mstr{s;s}\}]$&{\color{red}$[x/\{3,4\},\mathit{str}/\{\nil,\mstr{s},\mstr{s;s}\}]$}\\
\hline
\end{tabular}}
\end{center}
}
}
\caption{Some computations of the reflect executions in Example~\ref{es1}, with $s\defi x:=x+1$.}\label{fig:es2}
\end{figure*}
%
\end{example}

%
%

\section{Abstract Interpretation of Strings}\label{sect:asem} 

\subsection{The abstract domain}
Let $\Conf^\sharp = \CommS \times\Store^\sharp$ be the domain of abstract states, where $\ok{\Store^\sharp: \pc{\prog} \rarr{} \Mem^\sharp}$ denotes abstract stores ranged over $\ok{\astore}$, and $\ok{\Mem^\sharp:\Var \rarr{} \avalue}$ denotes the set of abstract memory maps ranged over by $\ok{\amem}$. The domain of abstract values for expressions is 
%
$\ok{\avalue\defi\{\top,\Inter,\Bool,\fa,\bot\}}$\footnote{Note that, we do not consider here implicit type conversion statements, namely each variable, during execution can have values of only one type, nevertheless we consider the reduced product of possible abstract values in order to define only one abstract domain.}. It is composed by $\Inter$, the standard GC-based abstract domain encoding the interval abstraction of $\wp(\mathbb{Z})$, by $\Bool$, the powerset domain of Boolean values, and by $\fa$ denoting the domain of FAs up to language equivalence. Given two FA $A_1$ and $A_2$ we have that $A_1 \equiv A_2$ iff $\cL(A_1) = \cL(A_2)$. Hence, the elements of the domain $\fa$ are the equivalence classes of FAs recognizing the same language ordered wrt language inclusion $\fa = \tuple{[A]_\equiv,\leq_\mathit{FA}}$, where $[A_1]_\equiv \leq_{\mathit{FA}} [A_2]_\equiv$ iff $\cL(A_1) \subseteq \cL(A_2)$. Here concretization is 
the language recognized $\cL$. By the Myhill-Nerode theorem \cite{bookComp} the domain is well defined and we can use the minimal automata to represent each equivalence class, moreover, the  ordering relation is well defined since it does not depend on the choice of the FA used to represent the equivalence class. In particular, we consider the domain $\fa$ defined over the finite alphabet $\Sigma$, thus, given $A \in \fa$, we have that $\cL(A) \in \wp(\Sigma^\ast)$.
%
FAs are closed for finite language intersection and union. They do not form a Galois connection with $\wp(\Sigma^\ast)$. 
%
The finite lub $\sqcup_\avalue$ and glb $\sqcap_\avalue$ among elements of $\avalue$ are defined as expected: the lub of two abstract values of the same type is given by the lub of the corresponding domain, while the lub between abstract values of different types is $\top$. This means that, for example, the lub of two intervals is the standard lub over intervals, while the lub between an interval and a FA is $\top$. Analogously, for the glb returning $\bot$ if applied to different types.  

Since $\Inter$ and $\fa$ are not ACC, and, in particular, $\fa$ is not closed by lubs of infinite chains, $\avalue$ is also not ACC and not closed. Therefore, we need to define a widening operator $\wid$ on $\avalue$. The widening operator among elements of different types returns $\top$, the widening operator of Boolean elements is the standard lub, $\Bool$ being ACC, the widening operator between elements of the interval domain is the standard widening operator on $\Inter$ \cite{CC79}. Finally, the widening operator on $\fa$ is defined in terms of the widening operator $\wid_R$ over finite automata introduced in \cite{silva-thesis}. 

Let us consider two FA $A_1= (Q^1,\delta^1,q_0^1,F^1,\Sigma^1)$ and $A_2=(Q^2,\delta^2,q_0^2,F^2,\Sigma^2)$ such that $\cL(A_1) \subseteq \cL(A_2)$: the widening between $A_1$ and $A_2$ is formalized  in terms of a relation $R \subseteq Q^1\times Q^2$ between the set of states of the two automata. The relation $R$ is used to define an equivalence relation $\equiv_R \subseteq Q^2 \times Q^2$ over the states of $A_2$, such that $\equiv_R = R \circ R^{-1}$. The widening between $A_1$ and $A_2$ is then given by the quotient automata of $A_2$ wrt the partition induced by $\equiv_R$: $A_1 \wid_R A_2 = A_2/\!\equiv_R$.
Thus, the widening operator merges the states of $A_2$ that are in equivalence relation $\equiv_R$.
By changing the relation $R$, we obtain different widening operators \cite{silva-thesis}. 
It has been proved that convergence is guaranteed when the relation $R_n \subseteq Q^1 \times Q^2$, such that $(q_1,q_2) \in R_n$ if $q_1$ and $q_1$, recognizes the same language of strings of length at most $n$~\cite{silva-thesis}. 
Thus, the parameter $n$ tunes the length of the strings determining the equivalence of states and therefore used for merging them in the widening. It is worth noting that, the smaller is $n$, the more information will be lost by widening automata. 
In the following, given two FA $A_1$ and $A_2$ with no constraints on the languages they recognize, we define the widening operator parametric on $n$ on $\fa$ as follows: $A_1 \wid_n A_2  \defi A_1 \wid_{R_n} (A_1 \sqcup A_2)$. %
%
%
%
\subsection{Abstract semantics of expressions}
\begin{figure}
\begin{center}
\includegraphics[scale=.]{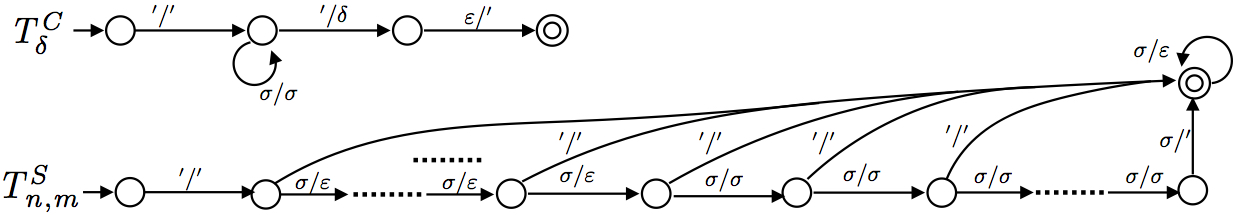}
\end{center}
\caption{SFT modeling string transformations with $\sigma\in\Sigma$.}\label{fig:string-sft}
\end{figure}

In this section, we model string operations, and in particular we observe that they can be expressed as SFTs, namely as symbolic transformers of a language of strings over $\Sigma$.  The SFTs that correspond to symbol concatenation $\Sexp\conc\sigma$ and to substring extraction $\subst{\Sexp}{\Aexp_1}{\Aexp_2}$  are given in Fig.~\ref{fig:string-sft}, where $\sigma$ ranges over the alphabet $\Sigma$.
In particular, for symbol concatenation we have an SFT $T^C_{\delta}$ for each symbol $\delta\in\Sigma$. Each SFT $T^C_\delta$ adds the considered symbol $\delta$ at the end of any string (note that if non deterministically we follow the $\varepsilon$ edge in the middle of a string then we cannot terminate in a final state anymore, meaning that the input string is not recognized and therefore no output is produced). As far as the sub-string operation is concerned, we have an SFT $T^S_{n,m}$ for each pair of non-negative values $n$ and $m$, which reads $n-1$ symbols in the input string without producing outputs, then it reads $m$ symbols from the $n$-th, releasing the symbol also in output, and finally it reads all the remaining symbols without producing outputs. It is clear that, if the string ends before reaching the starting point $n$, or before reading $m$ symbols, then the string is not accepted and no output is produced. Namely, if $\Sexp$ is the input string, the transformation works correctly only if $n+m\leq\length(\Sexp)$.
%
%

We can now define the abstract semantics of expressions as $\grasseb{\ExpS}^\sharp =\Mem^\sharp \rarr{}  \avalue$ as the best correct approximation of the collecting concrete semantics.
For instance, in Fig.~\ref{fig-aexpr} we specify the abstract semantics for string expressions. When we perform operations between expressions of the wrong type then we return $\top$, for example if we  add an interval to an FA. 

\subsection{Abstract Program Semantics}

We can now define the abstract transition relation $\rel^\sharp \subseteq \Conf^\sharp \times \Conf^\sharp$ among abstract states. The rules defining the abstract transition relation can be obtained from the rules of the concrete transition relation given in Fig.~\ref{fig:rules}, by replacing the collecting semantics of expressions $\grasseb{\cdot}$ with the abstract one,
and by modifying the assignment rules and the executability test. In particular, in the abstract transition, the memory update of the assignment rules uses the widening operator over $\avalue$ instead of the least upper bound. 
The executability test in the reflection rules is now $\cL(\grasseb{\Sexp}^\sharp\store_l^\sharp) \cap \ov{\CommS}$.
This allows us to compute the \emph{partial abstract collecting trace semantics} $\cF^\sharp(P,\cI^\sharp)$ of the abstract transition system $\tuple{\Conf^\sharp,\rel^\sharp}$.
Given a set of abstract initial states $\cI^\sharp \subseteq \Conf^\sharp$, we define the abstract fix-point function $F^\sharp:(\Conf^\sharp)^\ast \rarr{}\wp((\Conf^\sharp)^\ast)$, starting from $\cI^\sharp$, such that $\cF^\sharp(\P,\cI^\sharp) = \lfp(\mbox{$\mathtt{F}^\sharp_{\P,\cI^\sharp}$})$:
{\small \[
\begin{array}{ll}
\mathtt{F}^\sharp_{\P,\cI^\sharp}\defi\lambda X.\: \cI^\sharp\:\cup
                  \ssetf{\Statec_0^\sharp\ldots\Statec_i^\sharp\Statec_{i+1}^\sharp}{\Statec_0^\sharp\ldots\Statec_i^\sharp\in X,\ 
\Statec_i^\sharp\rel^\sharp\Statec_{i+1}^\sharp\!\!}
\end{array}
\]}
\begin{theorem}
$\cF^\sharp(\P,\cI^\sharp)$ is a sound approximation of $\cF(\P,\cI)$.
\end{theorem}
%
%
\begin{example}\label{rexe1}
Consider the following program fragment $P$
\begin{center}
{\footnotesize
\begin{tabular}{l}
$\pp{1}\whilec\ x<3$\\ $\qquad \{os:=os\conc '\!xA:=Bx+1B;y:=1A0;x:=Bx+1A;A\sep';\};$\\
$\pp{2}ds:=\textsl{deobf}(os);$\\
$\pp{3}\ifc\ x>10$\\ $\qquad \{os:='\!whiAleBx\!<\!5AA\{x:A=x+1;y:=x;\};B\sep';\};$\\
$\pp{4}ds:=\textsl{deobf}(os);$\\
$\pp{5}\ifc\ x=5\ \{os:='\!hello';\};$\\
$\pp{6}\ifc\ x=8\ \{os:='\!wBhilAeBx;';\};$\\
$\pp{7}ds:=\textsl{deobf}(os);$\\
$\pp{8}\reflect(ds);$\\
$\pp{9}\sep$
\end{tabular}}
\end{center}
where $ds:=\textsl{deobf}(os)$ is a syntactic sugar for the string transformer in Fig.~\ref{frexe1}
%
%
In Fig.~\ref{frexe1}-(a) is the FA, namely the abstract value, of $ds$ at program line $8$, computed by the proposed static analysis, wrt $\wid_3$.
\begin{figure}[ht]
\begin{center}
\includegraphics[scale=.25]{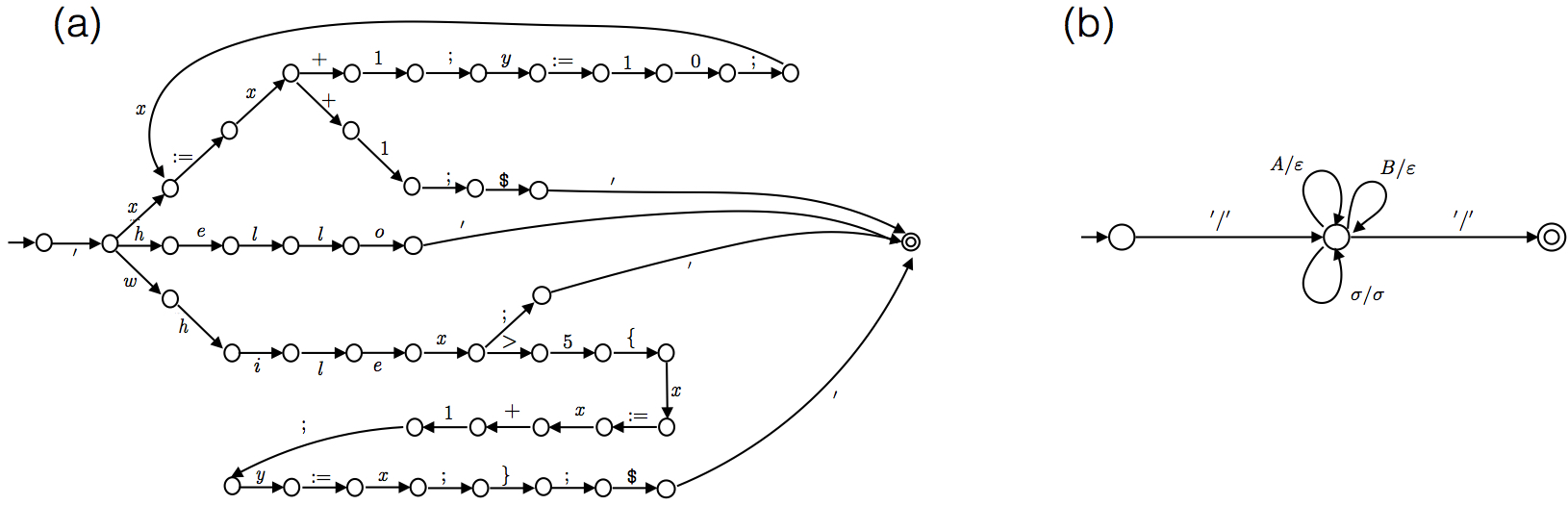}
\end{center}
\caption{FA $A^8_{ds}$ abstracting the value of $ds$ at program line $8$ of Ex.~\ref{rexe1}.}\label{frexe1}
\end{figure}
\end{example}
It is worth noting that, even in the approximate computation, we have the problem of decidability of the executability of $\grasseb{\Sexp}^\sharp\store_l^\sharp$. Indeed, it is still in general undecidable to compute the intersection $\cL(\grasseb{\Sexp}^\sharp\store_l^\sharp)\cap\ov{\CommS}$ between a possibly infinite language modeling the possible values of a string expression $\Sexp$ in a certain program point and a context free grammar (CFG) modeling the language. This means that, our implementation of the analysis needs to approximate the set of executable strings collected during the abstract computation for the arguments of reflection instructions.

\section{The SEA Analyzer}\label{approx2} 
The SEA analyser implements both a new string analysis domain and it performs an executability analysis in presence of a reflection statement.
SEA is indeed a prototype implementation with the ambition of providing a general language-independent sound-by-construction architecture for the static analysis of self modifying code, where only some components are language-dependent, in our case the abstract interpreter for $\CommS$.

\begin{figure}
\begin{center}
\includegraphics[scale=.55]{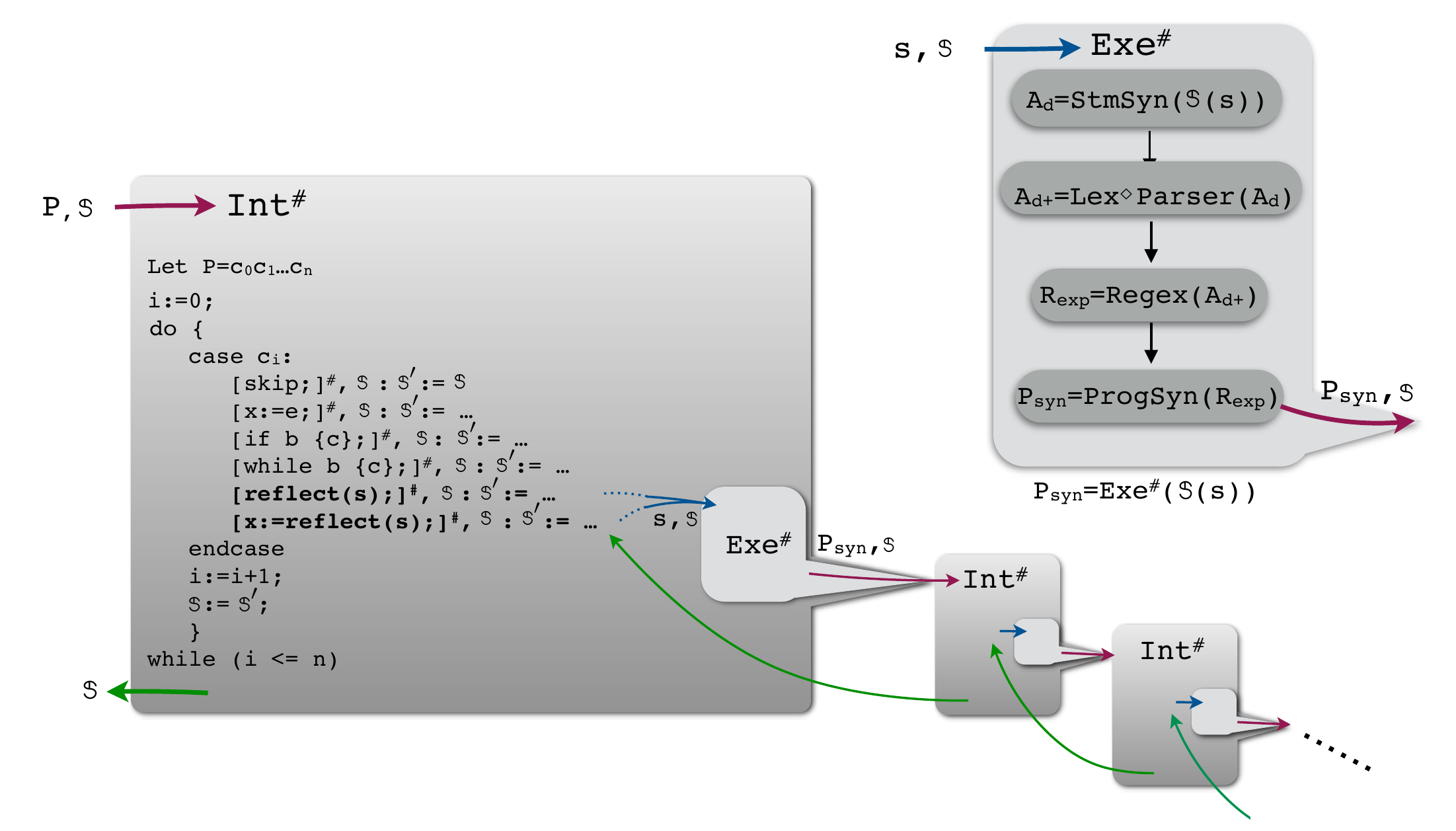}
\end{center}
\caption{Architecture and call execution structure of SEA.}\label{fig:archSEA}
\end{figure}

The first feature of SEA consists in the implementation of the interpreter based on the flow-sensitive collecting semantics proposed in Sect.~\ref{sect:asem}. The main original contribution is in the way the reflection analysis is handled. In particular, 
we provide an algorithmic approach for approximating in a decidable way the executability test $\cL(\grasseb{\Sexp}^\sharp\store^\sharp_l) \cap \ov{\CommS}$ and for building a program in $\CommS$ that soundly approximates the executable programs, i.e., whose semantics soundly approximates the semantics of the code that may be executed in a reflection statement. 
Our idea is first to filter the automaton collecting the string analysis in order to keep only an over-approximation of the executable strings and then to synthesise a code fragment whose possible executions over-approximate the possible concrete executions.
In Fig.~\ref{fig:archSEA} we show how SEA works, and we explain the architecture on a running example.
In Ex.~\ref{rexe1} we showed the execution of $\Int^{\#} (P,\store)$, where $\store$ starts with {\em any value} for $x$, up to program line $8$. Now, we can explain how the analysis works. In particular, following the execution structure in Fig.~\ref{fig:archSEA}, at line $8$ we call the execution of $\Exe^{\#}$ on $A^8_{ds}$ given in Fig.~\ref{frexe1} and in the following simply denoted $\ov{A}$.

The first step consists in reducing the number of states of the automaton, by over-approximating every string recognized as a statement, or partial statement, in $\CommS$.
%

\paragraph*{StmSyn.}
The idea is to consider the automaton computed by the collecting semantics $\mathtt{A}$, and to collapse all the consecutive edges up to any punctuation symbol in $\{\mbox{\small $;,\{,\},\sep$}\}$. In particular, any executable statement will end with $\mbox{\small $;$}$, while  $\mbox{\small $\{$}$ and $\mbox{\small $\}$}$ allow to split strings when the body of a $\whilec$ or of an $\ifc$ begins or ends, finally $\sep$ recognises the end of a program.
Hence, we design the  procedure $\textsc{Build}$, computed by Alg.~\ref{algoIq}, and recursively called by Alg.~\ref{algo} that returns 
an automaton on a finite subset of the alphabet:
$\Sigma_{\tt\tiny Syn}=\{\:\mbox{\small $\},\sep$}\}\cup\sset{x;}{x\in\Sigma^*}\cup\sset{x\mbox{\small $\{$}}{x\in\Sigma^*}$.
In particular, given the parsing tree $\mathtt{T_A}$ of the automaton $\mathtt{A}$, obtained by performing a depth first visit on $\mathtt{A}$, we define
\[\mathsf{Str}\defi\ssetf{x\in(\Sigma\smallsetminus\{\mbox{\small $;,\{,\},\sep$}\})^*}{\exists\ \mbox{path}\ \pi\ \mbox{in}\ \mathtt{T_A}\ \mbox{such that}\\ x\ \mbox{ maximal substring of}\ \pi}.
\]
Hence, the finite alphabet of the resulting automaton is
$\Sigma^\mathtt{A}_{\tt\tiny Syn}=\{\:\mbox{\small $\},\sep$}\}\cup\sset{x;}{x\in\mathsf{Str}}\cup\sset{x\mbox{\small $\{$}}{x\in\mathsf{Str}}$.
%
%
\begin{algorithm}
{\footnotesize\caption{Building the FA.}\label{algo}
\begin{algorithmic}[1]
\Require An FA $A=(Q,\delta,q_0,F,\Sigma)$
\Ensure  An FA $A'=(Q',\delta',q_0,F',\Sigma^\ast)$
\Procedure{\tt StmSyn}{$A$}
  \State $q_0'=\delta(q_0,')$ $//\mbox{The first apex \mbox{$'$} is erased}$
  \State $Q' \gets \{q_0'\};\ F'\gets F\cap\{q_0'\};\ \delta'\gets\varnothing$,\ Visited$\:\gets\{q_0'\}$; 
  \State \textsc{stmsyntr}$(q_0')$;
\EndProcedure
\Procedure{stmsyntr}{q}
  \State $B\gets$\textsc{Build}$(A,q)$;
  \State Visited $\gets$ Visited $\cup\{q\}$;\ $Q'\gets Q'\cup\sset{p}{(\ba,p)\in B}$;
  \State $F'\gets Q'\cap F$;\ $\delta'\gets \delta'\cup\sset{(q,\ba,p)}{(\ba,p)\in B}$;
  \State $W\gets\sset{p}{(\ba,p)\in B}\smallsetminus$Visited; 
  \While{$W\neq\varnothing$}
    \State select $p$ in $W$ ($W\gets W\smallsetminus\{p\}$);
    \State \textsc{stmsyntr}$(p)$;
  \EndWhile
\EndProcedure
\end{algorithmic}}
\end{algorithm}
%
%

\begin{algorithm}
{\footnotesize\caption{Statements recognized from a state $q$.}\label{algoIq}
\begin{algorithmic}[1]
\Require An FA $A=(Q,\delta,q_0,F,\Sigma)$
\Ensure  $I_q$ set of all pairs (statement,reached state)
\Procedure{Build}{$A,q$}
  \State $I_q \gets \varnothing$ 
  \State $\textsc{buildtr}$(q,$\varepsilon$,$\varnothing$) 
\EndProcedure
\Procedure{buildtr}{q,word,Mark}
  \State $\Delta_q\gets\sset{(\sigma,p)}{\delta(q,\sigma)=p}$  
  \While{$\Delta_q\neq\varnothing$}
    \State select $(\sigma,p)$ in $\Delta_q$ 
    ($\Delta_q\gets\Delta_q\smallsetminus\{(\sigma,p)\}$)
     \If{$(q,p)\notin$ Mark}
        \If{$\sigma\notin\{\mbox{\small $;,\{,\},\sep$}\}\ \wedge\ p\notin F$}
        	\State $\textsc{buildtr}$(p,word.$\sigma$,Mark$\cup\{$(q,p)$\}$)
        \EndIf	
  		\If{$\sigma\in\{\mbox{\small $;,\{,\},\sep$}\}$}
  			 $I_q \gets I_q\cup\{($word$.\sigma,p)\}$  
  		\EndIf
  		\If{$\sigma=\:'\ \wedge\ p\in F$} $I_q \gets I_q\cup\{($word$,p)\}$  
  		\EndIf    
      \EndIf
  \EndWhile
\EndProcedure
\end{algorithmic}}
\end{algorithm}
%
%

The idea of the algorithm is first to reach $q_0'$ from $q_0$ reading the symbol $'$, and then
to perform, starting from $q_0'$, a visit of the states recursively identified by Algorithm~\ref{algoIq} and to recursively replace the sequences of edges that recognize a symbol in $\Sigma^\mathtt{A}_{\tt\tiny Syn}$ with a single edge labeled by the corresponding string. In particular, from $q_0'$ we reach the states computed by $\textsc{Build}(q_0')$, and the corresponding read words. Recursively, we apply $\textsc{Build}$ to these states, following only those edges that we have not already visited.
It is clear that, in this phase all the non-executable strings not ending with a symbol in $\{\mbox{\small $;,\{,\},\sep$}\}$ are erased from the automata, hence we have a reduction of non executable strings.
For instance, in Fig.~\ref{frexe5} we have the computation of $\texttt{StmSyn}(\ov{A})$, denoted $\ov{A}_{\mbox{\tt\tiny d}}$.  
\begin{figure}[ht]
\begin{center}
\includegraphics[scale=.35]{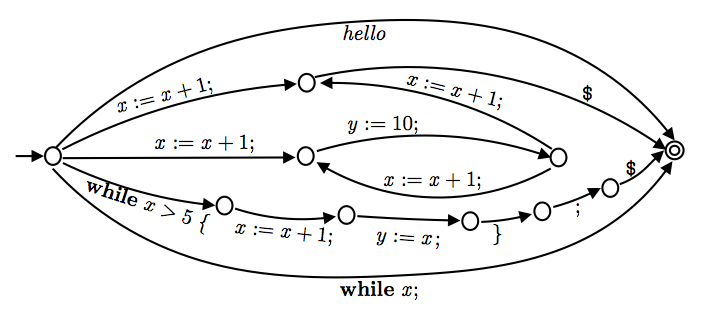}  
\end{center}
\caption{Automaton $\ov{A}_{\mbox{\tt\tiny d}}=\texttt{StmSyn}(\ov{A})$.}\label{frexe5}
\end{figure}
From the computational point of view, 
we can observe that the procedure $\textsc{Build}(A,q)$ executes a number of recursive-call sequences equal to the number of maximal acyclic paths starting from $q$ on $A$. The number of these paths can be computed as $\sum_{q \in Q} (\mathit{outDegree}(q) - 1) + 1$, where $\mathit{outDegree}(q)$ is the number of outgoing edges from $q$. The worst case depth of a recursive-call sequence
is $|Q|$. Thus, the worst case complexity of $\textsc{Build}$ (when $\mathit{outDegree}(q)=|Q| \times |\Sigma|$ for all $q \in Q$)  is $O(|Q|^3)$.
As far as $\texttt{StmSyn}$ is concerned, we can observe that in the worst case we keep in $\texttt{StmSyn(A)}$ all the $|Q|$ states of $A$, hence in this case we launch $|Q|$ times the procedure $\textsc{Build}$, and therefore the worst case complexity of $\texttt{StmSyn}$ is $O(|Q|^4)$.
\\
Next step consists in verifying whether the labels of each edge in $\texttt{StmSyn}(\mathtt{A})$ are potentially executable, or portion of an executable statement.

\paragraph*{Lex-Parser.}
In order to proceed with the analysis, we need to synthesize a program from $\mbox{\tt A}_{\mbox{\tt\tiny d+}}$
approximating the set of executable string values assumed by string $\Sexp$ at program line $l$ where reflection is executed. This would allow us to replace the argument of the reflect with the synthesized program and use the same analyser (abstract interpreter) for the analysis of the generated code. 
Hence, we have to check whether each label in $\mbox{\tt A}_{\mbox{\tt\tiny d}}=\texttt{StmSyn}(\mathtt{A})$ is in particular in the
alphabet $\CommS^- \subseteq \Sigma^\mathtt{A}_{\tt\tiny Syn}$ of (partial) statements of $\CommS$ statements:

\[
\CommS^-\defi \set{ \skipc;, x:=\Exp;, \textbf{if} ~\Bexp~\{, \textbf{while} ~\Bexp~ \{,\\ \reflect(\Sexp);, x := \reflect(\Sexp);,\},\sep}
\]
where $\Sexp,\Exp,\Bexp$ are expressions in the language $\CommS$. Hence, we need a parser for the language $\CommS^-$. This parser can me modelled as the composition of two SFTs. The first one, $\lex$, has to recognises the lexemes in the language by identifying the language tokens. We consider the following set of tokens for $\CommS^-$. These tokens correspond to the terminals of $\CommS^-$ except for the punctuation symbols $\pun \defi \{\mbox{\small $;,\{,\},),\sep$}\}$ that will be directly handled by the parser.

\vspace{-.2cm}
\[
\token\defi\set{\texttt{id},\texttt{const}_{\Sexp},\texttt{const}_{\Aexp},\texttt{const}_{\Bexp},\texttt{aop},\texttt{bop},\texttt{uop},\\\texttt{num},\texttt{len},\texttt{conc}_\delta,\texttt{substr},\texttt{relop},\texttt{if},\texttt{while},\\\texttt{assign},\texttt{skip},\texttt{reflect},\texttt{rand}}
\]
For each token $\texttt{T} \in \token$, it is possible to define an SFT that recognises its possible lexemes and outputs the lexemes followed by the token name. Let us denote with $T_\texttt{T}$ the SFT that recognises the lexemes of the token $\texttt{T} \in\token$, so, for example, $T_{\texttt{id}}$ is the SFT corresponding to the token $\texttt{id}$.
The transduction is $\fT_{\lex}:\Sigma^\ast\lra (\Sigma \cup  \token)^\ast$ defined as:

\vspace{-.2cm}
\begin{multline*}
\fT_{\lex}(\ba) \defi
\left \{
\begin{array}{ll}
\ba^0 \texttt{T}^0 \bp^0 \ba^1 \texttt{T}^1 \bp^1 \dots \ba^n\texttt{T}^n \bp^n& \mbox{if }\ba = \ba^0 \cdot \ba^1 \cdot ... \cdot \ba^n \in \Sigma^\ast\\
& \forall i \in [0,n]: \, \ba^i \in \cL(T_{\texttt{T}^i}),\\
& \texttt{T}_i \in \token, \bp^i \in \pun^\ast\\
\emptyset & \mbox{otherwise}\\
\end{array}  
\right.
\end{multline*}

In order to build the $\parser$, we design also the SFT recognising the correct sequences of lexemes and tokens that build respectively arithmetic, boolean and string expressions, and which correctly combines them in order to obtain objects in the language $\CommS^-$.
Hence, $\parser$ should implement the transduction function $\fT_{\parser}: (\Sigma \cup  \token)^\ast \ra \Sigma^*$ is such that:
$\ba \in \CommS^- \; \Ra \; \fT_{\parser}(\fT_{\lex}(\ba)) = \ba$.
This means that the composition $\lex \diamond \parser$ allows sequences of $\Sigma^\mathtt{A}_{\tt\tiny Syn}$ which are in $\CommS^-$. The other implication does not hold since $\parser$ allows also sequences of commands of $\CommS^-$ that contain syntactic errors due an erroneous number of punctuation symbols in $\pun$.  This means that for example the sequence $\mstr{x\texttt{id}:= \texttt{assign}x\texttt{id}+\texttt{aop}1\texttt{const$_{\mbox{\tt a}}$};;;\mathit{skip}\texttt{skip}}$ is allowed by $\parser$ and given in output as it is. 
In Fig.~\ref{frexe3} we can find the automaton $\ov{A}_{\mbox{\tt\tiny d+}}$, which is $\ov{A}_{\mbox{\tt\tiny d+}}$ where all the sequences which are not in $\CommS^-$ are erased.

This module is implemented in SEA using JavaCC \cite{jcclink}: given in input a BNF-style definition of a grammar $G$ it returns as output the parser for $G$.

\begin{figure}[t]
\begin{center}
\includegraphics[scale=.4]{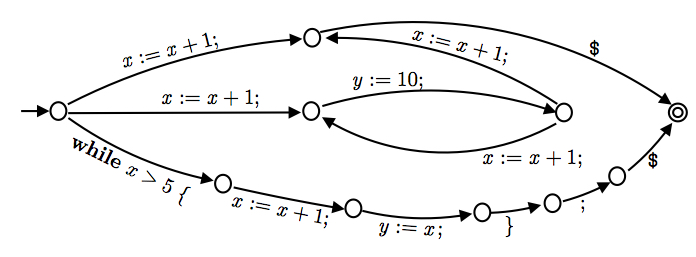}
\end{center}
\caption{
Executable automaton $\ov{A}_{\mbox{\tt\tiny d+}}=\lex \diamond\parser(\ov{A}_{\mbox{\tt\tiny d}})$.}\label{frexe3}
\end{figure}

\paragraph*{Regex.}
The so far obtained automaton can be used to synthesize a program by extracting the regular expression corresponding to the language it recognizes \cite{oz64}. Let $\textsl{RE}$ be the domain of regular expressions over  $\CommS^-$, and $\regf:\mbox{FA}\ra\textsl{RE}$ be such an extractor. For instance, in the running example, $\ov{R}_{\mbox{\tt\tiny exp}}=\regf(\ov{A}_{\mbox{\tt\tiny d+}})$ is the following regular expression (with standard operators in boldface):
\begin{center}
{\footnotesize
\begin{tabular}{l}
$\ov{R}_{\mbox{\tt\tiny exp}}=\ x:=x+1;\sep\:${\bf\large +}$\:\whilec\:x>5\:\{x:=x+1;y:=x;\};\sep$\\
{\bf +}$\:x:=x+1;y:=10;${\bf\large (}$x:=x+1;y:=10;${\bf\large )}$^*x:=x+1;\sep\:$
\end{tabular}}
\end{center}
SEA implements the Brzozowski algebraic method~\cite{oz64} to convert an automaton to an equivalent regular expression. 

\paragraph*{ProgSyn.}
Finally, we define $\mathtt{ProgSyn}$ implementing the function $\trad{\cdot}_{\P}: \textsl{RE} \ra \CommS$ that, given a regular expression $\Rexp \in \textsl{RE}$, translates it into a program in $\CommS$. This is defined in terms of a translation function $\trad{\cdot}:\textsl{RE}\ra\Ccomms$ (erasing $\sep$) inductively defined on the structure of the regular expression $\Rexp$: Let us denote by $\Dexp_;$ the symbol $\Dexp$ without the last $;$ (e.g., $(x:=x+1;)_;=x:=x+1$)

{\footnotesize
\[
\begin{array}{rl}
\trad{\Dexp} = & \Dexp_; \mbox{ if } \Dexp \in \CommS^-\\
\trad{\Rexp\sep} = &\trad{\Rexp};\\
\vspace{.1cm}
\trad{\Rexp_1\Rexp_2} = & \trad{\Rexp_1\!};\trad{\Rexp_2}; \\
\vspace{.1cm}
\trad{\textbf{\large (}\Rexp\textbf{\large )}^*} = &
\left [
\begin{array}{l}
g: = \rand();\\
\whilec\ g = 1\ \{\trad{\Rexp};g: = \rand();\};
\end{array}
\right.\\
\vspace{.0cm}
\trad{\Rexp_1\textbf{\large +}\Rexp_2} = & 
\left [
\begin{array}{l}
g: = \rand();\\
\textbf{if } g = 1 \, \{\trad{\Rexp_1};\}; \textbf{if } g = 2 \, \{\trad{\Rexp_2};\};
\end{array}
\right.
\end{array}
\]}
and $\trad{\Rexp}_{\P}=\labb{\trad{\Rexp}\sep}$.
Hence, in our running example, the synthesis from the regular expression $\ov{R}_{\mbox{\tt\tiny exp}}$, i.e., $\ov{P}_{\mbox{\tt\tiny syn}}=\mathtt{ProgSyn}(\ov{R}_{\mbox{\tt\tiny exp}})$, is the program\\
\\
{\footnotesize
\begin{tabular}{l}
$\pp{1}g1:= \rand();$\\
$\pp{2}\ifc\ g1=1\ \{\pp{3}x:=x+1;\};$\\
$\pp{4}\ifc\ g1=2\ \{$\\
\qquad$\pp{5}g2:=\rand();$\\
\qquad$\pp{6}\ifc\ g2=1\ \{\pp{7}\whilec\ x>5\ \{\pp{8}x:=x+1;\pp{9}y:=x;\};\};$\\
\qquad$\pp{10}\ifc\ g2=2\ \{$\\
\qquad\qquad$\pp{11}x:=x+1;\pp{12}y:=10;\pp{13}g3=\rand();$\\
\qquad\qquad$\pp{14}\whilec\ g3 = 1\ \{\pp{14}x:=x+1;\pp{15}y:=10;$\\
\qquad\qquad\qquad\qquad\qquad\ \ \ \ \ \: $\pp{16}g3=\rand();\};$\\
\qquad\qquad$\pp{17}x:=x+1;\};$\\
\qquad$\};\pp{18}\sep$
\end{tabular}}

\paragraph*{Soundness.}
Next theorem proves the soundness of the approximate program synthesis. Safety (i.e., prefix closed) properties of dynamically generated code are soundly approximated by the synthesized program output of our analysis. 

\begin{theorem}\label{th}
Let $\P\in\CommS$ containing $\pp{l}\reflect(\Sexp)$, and let $\store\in\Store$ be the store on which $\P$ is executed. Then, for any $\store'\in\Store$ such that $\store'_1=\store_l$ the partial semantics  of any statement in the evaluation of $\Sexp$ executed from $\store'$ is contained the partial semantics of the synthesized program, formally
{\small
\[
\begin{array}{l}
\forall\Comm\in \grasseb{\Sexp} \store_l\cap\ov{\CommS}.\:\\
\grasstr{\Comm}\store'\subseteq 
\grasstr{\mathtt{ProgSyn}(\mathtt{Regex}(\mathtt{Lex}\diamond\mathtt{Parser}(\mathtt{StmSyn}(\mathtt{A}_{\Sexp}^l))))}\store'.
\end{array}
\]}
\end{theorem}
%

\paragraph*{Termination.}
As observed in Example~\ref{infTower}, the use of reflection suffers from the potential divergence of unbounded nested reflection which goes beyond the control of widening. In this case, the divergence comes directly from the meaning of reflection and cannot be controlled by the semantics once we execute the reflect statement, hence also our analysis in this situation would diverge. SEA ensures soundness until a maximal degree of nested calls to $\reflect$.

In order to keep soundness beyond a maximal degree of nested reflections, we can introduce a widening with threshold, i.e., the widening acts after a given number of calls to the abstract interpreter. This corresponds to fix a maximal allowed height of towers, fixing the degree of precision in observing the nesting of reflect statements. Given a \emph{tower height threshold} $\tilde{\tau}$ such that, any tower higher than $\tilde{\tau}$ is approximated as computing any possible value for the program variables whose name is a substring of the string evaluated at $\tilde{\tau}$, therefore guaranteeing soundness. In order to check the height of towers, we need to enrich the store by including a new numerical variable $\tau$ counting the nesting level of reflection. Let $\Store^{\tilde{\tau}}:\Loc\ra \Mem^{\tilde{\tau}}$ be this enriched domain, where $\Mem^{\tilde{\tau}}:\Var\cup\{\tau\}\ra\wp(\mathbb{Z})\cup\Bool\cup\wp(\Sigma^*)\cup\mathbb{Z}$. Hence, we can define a new 
partial trace semantics $\grasstr{\cdot}^{\tilde{\tau}}$ on the transition system $\tuple{\Conf^{\tilde{\tau}},\rel^{\tilde{\tau}}}$ associated with programs in $\CommS$, where $\Conf^{\tilde{\tau}}=\CommS\times\Store^{\tilde{\tau}}$ is the set of states in the transition system and $\rel^{\tilde{\tau}}\subseteq\Conf^{\tilde{\tau}}\times\wp(\Conf^{\tilde{\tau}})$ is the transition relation. Note that, the semantics of all statements does not change (supposing that $\mem_\varnothing\in\Mem^{\tilde{\tau}}$ associates $0$ to $\tau$) except for $\reflect$, whose new 
rule should count the number of recursive interpreter $\Int^{\#}$ calls.

\section{Evaluation}\label{evaluation} 
\begin{figure*}
\scalebox{0.62}{%
\vbox{%
\hspace{1.5cm}
\begin{tabular}{|l|l|l|}
\hline
 P & TAJS analysis of $y$ & TAJS reflection of $y$ \\
\hline\hline
\begin{tabular}{l}
$y:=\mstr{x=x+1;};\reflect(y);$
\end{tabular}
&$\mstr{x=x+1;}$&$x:=x+1;$\\
\hline
\begin{tabular}{l}
$\ifc\ x>0 \{y:=\mstr{a:=a+1;}\}; $\\
$\ifc\ x<0 \{y:=\mstr{b:=b+1;}\}; \reflect(y);$\\
\\
\end{tabular}
&$\mathtt{String}$&$\mathtt{AnalysisLimitationException}$\\
\hline
\begin{tabular}{l}
$x:=1; \mathit{y}:= \nil;$\\
$\whilec \; x < 3 \; \{\mathit{y} := \mathit{y} \conc \mstr{x:=x+1;}; x:=x+1;\};$\\
$\mathit{y}:=\mathit{y}\conc\mstr{\$};\reflect (\mathit{y});$\\
\\
\end{tabular}&$\mathtt{String}$&$\mathtt{AnalysisLimitationException}$\\
\hline
\end{tabular}
%

\begin{center}
\begin{tabular}{|l|l|l|}
\hline
 P & SEA analysis of $y$ & SEA reflection of $y$\\
\hline\hline
\begin{tabular}{l}
$y:=\mstr{x=x+1;};\reflect(y);$
\end{tabular}
& \includegraphics[scale=.3]{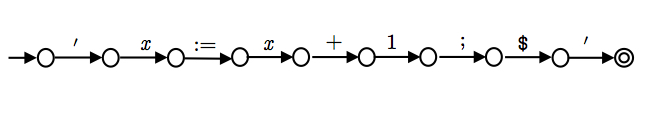}&\begin{tabular}{l}
$x:=x+1;\sep$\\
\\
\end{tabular}\\
\hline
\begin{tabular}{l}
$\ifc\ x>0 \{y:=\mstr{a:=a+1;}\}; $\\
$\ifc\ x<0 \{y:=\mstr{b:=b+1;}\}; \reflect(y);$\\
\\
\end{tabular}
&\includegraphics[scale=.3]{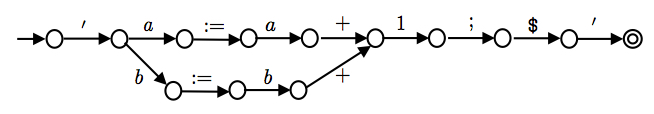}&
\begin{tabular}{l}
$g1 := \rand ();$\\
$\ifc\ g1 = 1 \{a := a +1;\};$\\
$\ifc\ g1 = 2 \{b := b +1;\};\sep$\\
\\
\end{tabular}\\
\hline
\begin{tabular}{l}
$x:=1; \mathit{y}:= \nil;$\\
$\whilec \; x < 3 \; \{\mathit{y} := \mathit{y} \conc \mstr{x:=x+1;}; x:=x+1;\};$\\
$\mathit{y}:=\mathit{y}\conc\mstr{\$};\reflect (\mathit{y});$\\
\\
\end{tabular}&\includegraphics[scale=.3]{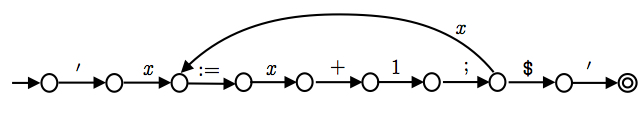} & \begin{tabular}{l}
$x:=x + 1;g1:=\rand();$\\ 
$\whilec\  g1 = 1\; \{$\\
$\quad x:=x + 1\; g1:=\rand(); \};\sep$\\
\\
\end{tabular}\\
\hline
\end{tabular}
\end{center}
}
}
\caption{SEA \textit{vs} TAJS}\label{confronto}
\end{figure*}
\begin{figure*}[h]
\begin{center}
\includegraphics[scale=.6]{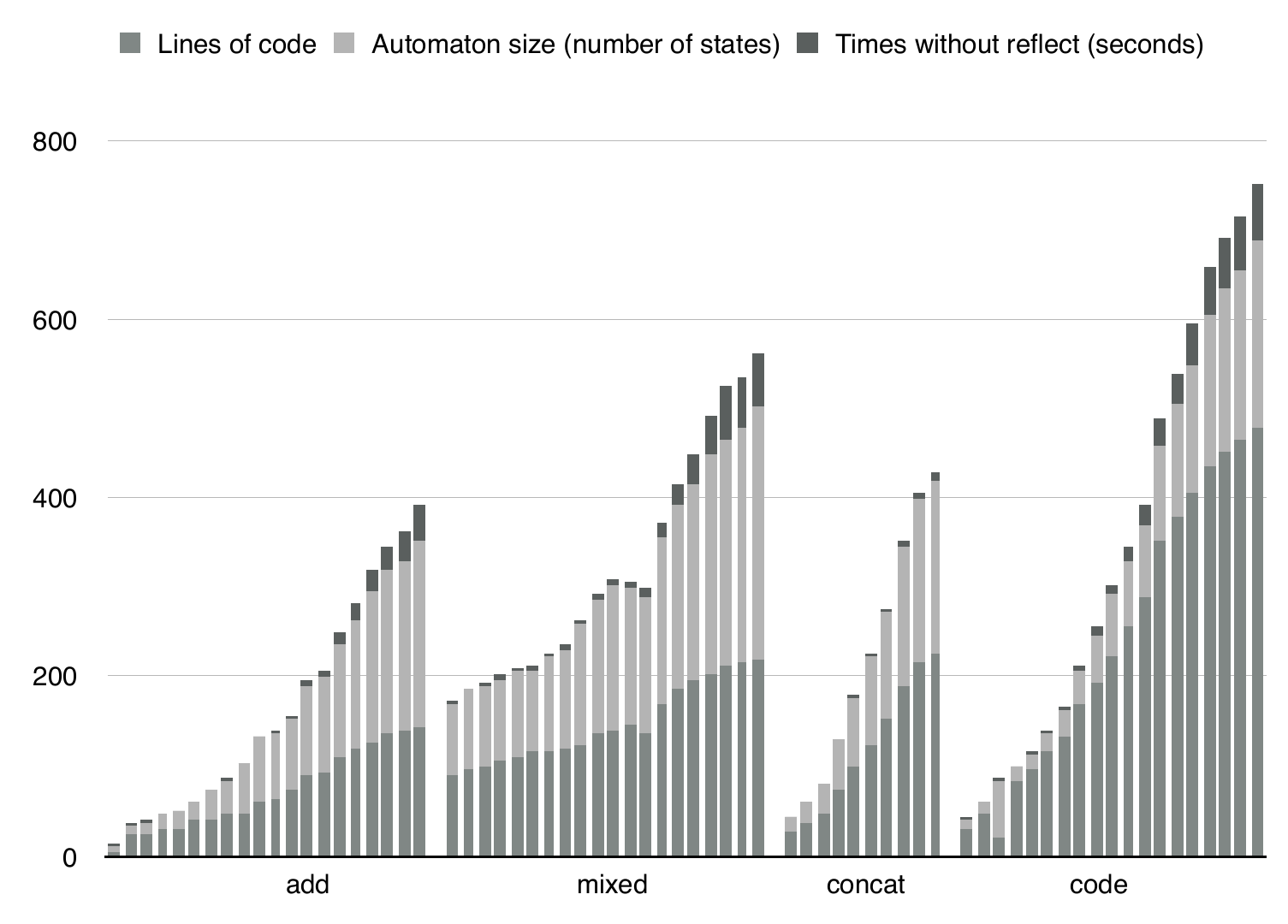}\hspace{2cm}
\end{center}

\begin{center}
\includegraphics[scale=.6]{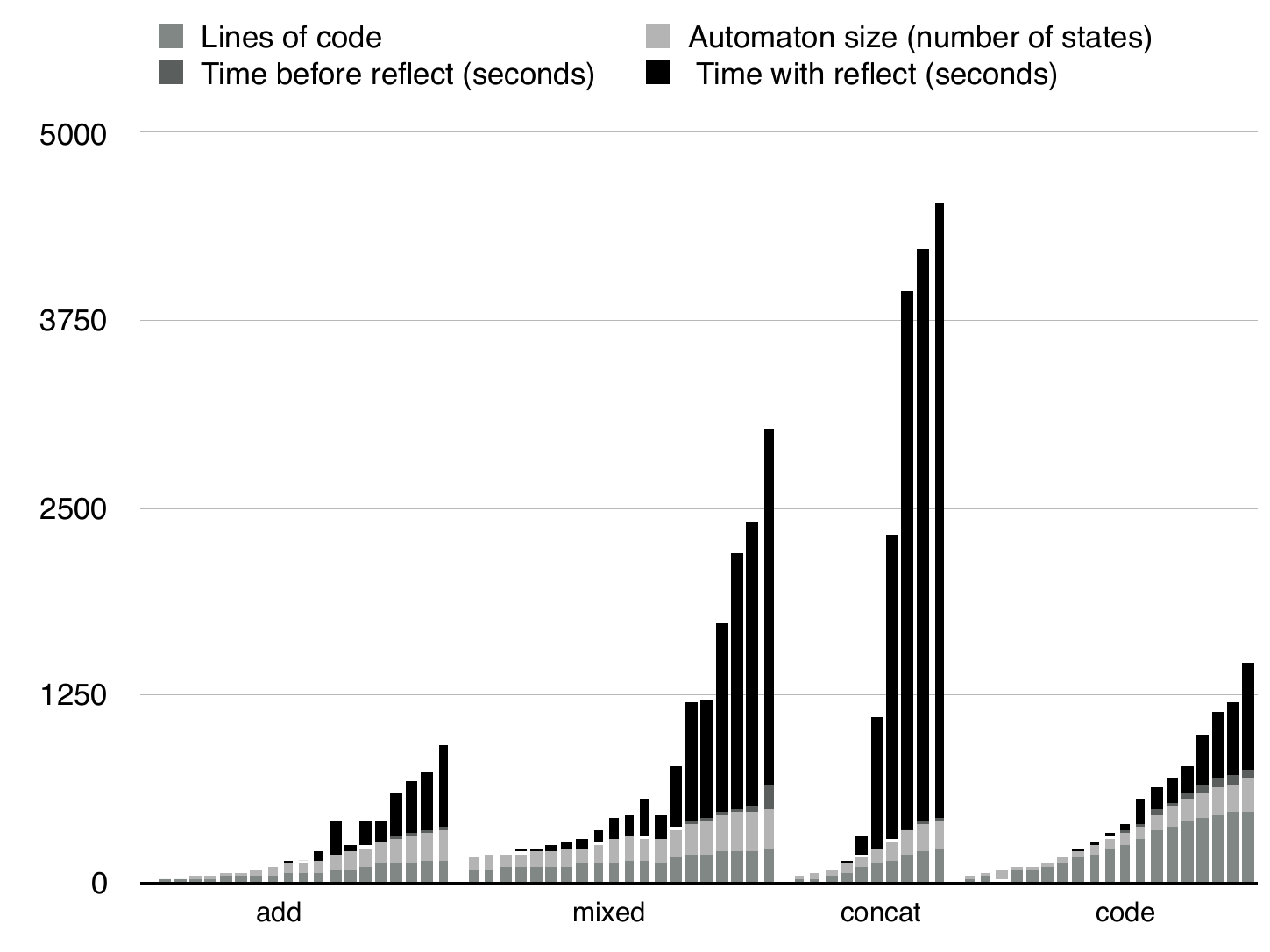}
\end{center}
\caption{Execution times in secs.\ without reflection (top) and with reflection (bottom). We ran the tests on an Intel i5-4210u 2.20 GHz processor.}\label{fig:no-ref}\label{fig:ref}
\end{figure*}

SEA is a proof of concept, showing that it is possible to design and implement an efficient sound-by-construction static analyser based on abstract interpretation for self modifying code written in high-level script-like languages. It was not in the intention of SEA to be optimal and directly applicable to existing script dynamic languages, such as PHP or JavaScript. We implemented SEA in Java 1.8 and we tested it on some significant code examples in order to highlight the strengths and the weaknesses of the analyser presented. In particular, we evaluate the precision of our string abstract domain as compared to TAJS~\cite{tajs09,tajs-talk,tajs-tool} (version 0.9-8), which is one of the best static analyser available for JavaScript based on abstract interpretation. To the best of our knowledge TAJS is the only tool statically analysing string-to-code primitives such as \textbf{eval}. This approach basically consist of a sound transformation of a JavaScript program $P_{\mbox{\tt\tiny eval}}$, containing \textbf{eval}, in another JavaScript program $P_{\mbox{\tt\tiny uneval}}$ where the \textbf{eval} statement is substituted with its argument, obviously converted in executable code, when this is possible, namely when the code to execute can be statically extracted as a constant form $P_{\mbox{\tt\tiny eval}}$.
All examples in the next sections have been compiled from $\CommS$ into a semantics-equivalent JavaScript program, in order to perform the comparison with TAJS. 

\subsection{Precision}\label{precision}

We performed more than 100 tests on programs of variable length, 70 of them are used to test the SEA performances and will be addressed in Sect.~\ref{perf}.
~We observed that the results can be classified in three different classes depending on some features of the analysed program.
We report three significant examples in Fig.~\ref{confronto} where we also compare SEA with TAJS.

The first class of tests consists in all the programs where the string variables collect only one value during execution, i.e., they are constant string variables. A toy example in this class of programs is provided in the first row of Fig.~\ref{confronto}, where the string value contained in $y$ is hard-coded and constant. In this case, both SEA and TAJS are precise and no loss of information occurs during the analyses.
By using the value of $y$ in SEA as input of $\reflect$, we obtain exactly the statement $x:=x+1;$ since $\mathtt{Exe}^\sharp$ behaves as the identity function. TAJS performs the uneval transformation and executes the same statement.

The second class of tests consists in all the programs where there are no constant string variables, namely variables whose value before the reflection is not precisely known being a set of potential string values. 
As toy example of this class of programs, consider the snippet of code in the second row of Fig.~\ref{confronto}, a simplification of Example \ref{rexe1}. In this case, since we don't have any information about $x$ we must consider both the branches, which means that before the reflection we only know that $y$ is one value between $'a:=a+1'$ and $'b:=b+1'$. 

If we analyse this program in TAJS, we observe that, after the if statement, the value of $y$ is identified as a string, since TAJS does not perform a collecting semantics and when it loses the constant information it loses the whole value. Unfortunately, this loss of precision, in the analysis of $y$, makes the TAJS analysis stuck, producing an exception when the reflection statement is called on the non constant variable. 
On the other hand, SEA computes the collecting semantics and therefore it keeps the least upper bound of the stores computed in each branch, obtaining the abstract value for $y$ modelled by the automaton $A_y$ in the second row equivalent to the regular expression $\mstr{a:=a+1;} + \mstr{b:=b+1;}$. Afterwards, the SEA analyser returns and analyse the sound approximation of the program passed to the reflection statement reported in the second row, which is the result of $\Exe^{\#}(A_y)$.

In the last class of examples, the string that will be executed is not constant and it is dynamically built during execution. In the simple example provided in Fig.~\ref{confronto} the dynamically generated statement is $\mstr{x:=x+1; x:=x+1;}$. In this case, as it happened before, TAJS loses the value of $y$ (which is a set of potential strings) and can only identify $y$ as a string. This means that, again, the reflection statement makes the analysis stuck, throwing an exception.
On the other hand, SEA performs a sound over-approximation of the set of values computed in $y$. In particular, the analysis, in order to guarantee termination and therefore decidability, computes widening instead of least upper bound between automata inside the loop. This clearly introduces imprecision, since it makes us lose the control on the number of iterations. In particular, instead of computing the precise automaton containing only and all the possible string values for $y$ (as in the previous case) we compute an automaton strictly containing the concrete set of possible string values. The computed automaton is reported in the third row and it is equivalent to the regular expression $\mbox{$\:'x:=x+1;(x:=x+1;)^*\;'$}$. The presence of possible infinite sequences of $\mstr{x:=x+1;}$ is due to the over-approximation induced by the widening operator $\wid_3$ on automata. Nevertheless, note that the widening parameter can be chosen by the user in order to tune the desired precision degree of the analysis: the higher is the parameter the more precise and costly is the analysis. It should be clear that, the introduced loss of precision increases the imprecision in any further analyses which uses the synthesised code. 
The synthesis of the program from the abstract value of $y$ returns the code reported in the third row: due to widening, as observed above, the command $\mstr{x:=x+1}$ can diverges.

A final observation on precision concerns the analysis of programs with unknown inputs. SEA considers an unknown input as a variable that may assume any possible string value. It is clear that in this kind of situations, TAJS necessarily get stuck whenever something depending on this unknown input is executed. Instead, SEA can keep some information since the abstract value consisting in {\em any possible value} is modelled by the automaton recognising $\Sigma^*$. In this way SEA can trace the string manipulations (substring and concatenation) on the unknown input, occurring during the execution.
\subsection{Performances}\label{perf}
We have tested the performances of SEA on a benchmark of 70 increasingly complex programs. Each program manipulates an automaton and finally it reflects the string value. 
The benchmark can be clustered in four families depending on the kind of string operations considered in the programs, determining the kind of automata manipulations performed by the analysis: \textit{\textbf{add}} (programs where the manipulation of strings adds always new whole statements, this corresponds, in our analysis, to adding completely new paths in the automaton), \textit{\textbf{concat}} (programs where the manipulation of strings concatenates new paths to those in the automaton),
 \textit{\textbf{mixed}} (programs performing both the manipulations), 
 \textit{\textbf{code}} (programs in the \textit{\textbf{add}} family, where we have added statements not manipulating strings, i.e., standard code not affecting strings).
 
%
%
Fig.~\ref{fig:no-ref}-top, shows the results obtained from the benchmark concerning the string analysis without reflection: increasing the lines of code as well as the number of the automaton states, the execution time increases with an almost linear trend.

In Fig~\ref{fig:ref}-bottom, we show the results due to string executability analysis. 
The total execution time increases more quickly of both than the length code and the automata dimension. But, it is worth noting that most of the time increase is due to the execution of the code generated by $\Exe^{\#}$ (the top black portion of the bars in Fig.~\ref{fig:no-ref}-bottom), while the time cost of the analysis still increases with an almost linear trend. This outcome tells us that the string and executability analysis scale quite well on the source original code, but it gets worse on the code generated by $\Exe^{\#}$. We believe that this is due to the implementation of the synthesised code, which does not optimise the code generated by $\Exe^{\#}$. The optimisation of the generated code is a future work that deserves further investigation.
%

\subsection{Analysis limitations and conclusions}
SEA attacks an extremely hard problem in static program analysis, providing the very first proof of concept in sound static analysis for self-modifying code based on bounded reflection in a high-level script-like language.
The main limitations of SEA are two: the simplicity of the programming language analysed, missing some important language features such as procedure calls and objects, and the fact that $\CommS$ is not a real script language.
The choice of keeping the language as simple as possible is due to the aim of designing a core analyser focusing mainly on string executability analysis.
We are conscious that, in order to implement a real-world static analyser we will have to integrate in our language several more sophisticated language features, but this is beyond this proof of concept. For instance, an extension would be the possibility of allowing implicit type conversion statements provided by many modern languages, such as PHP, JavaScript or Python.

As far as the second limitation is concerned, we already observed that this choice is due to the ambition of providing the most general possible architecture for executability string analysis. We believe that SEA is a step towards an implementation of a sound-by-construction analyser for reflection in real dynamic languages, since its design is fully language independent. In particular, the SEA architecture is invariant on the choice of the string abstraction, in our case FA, as well as the other state abstractions, in our case intervals, and on the language features, provided that their formal semantics is given.

%
%
%
%
%
%



\bibliographystyle{acm}
\bibliography{Biblio}

\end{document}